\documentclass[10pt,journal,twocolumn]{IEEEtran}
\usepackage{amsthm}
\usepackage{amssymb}
\usepackage{amsfonts}
\usepackage{amsfonts,subfigure,multicol,color,verbatim,cite,epsfig,amssymb,amsmath,cases,bm,xcolor,multirow,array}
\usepackage{graphicx}
\usepackage{epstopdf}
\usepackage{algorithm,algorithmicx,algpseudocode}
\usepackage{multirow}
\usepackage{multicol}
\usepackage{supertabular}
\usepackage{array}
\usepackage{colortbl}
\usepackage{longtable}
\usepackage{subfigure}
\usepackage{dsfont}
\usepackage{mathrsfs}
\usepackage{CJK}
\usepackage{indentfirst}
\usepackage{booktabs}
\usepackage{color}
\usepackage{cuted}\stripsep -3pt plus 3pt minus 2pt

\newcounter{MYtempeqncnt}
\begin{document}
\title{Distributed Satellite-Terrestrial Cooperative Routing Strategy Based on Minimum Hop-Count Analysis in Mega LEO Satellite Constellation}
\author{{Xin'ao~Feng,
             Yaohua~Sun,
             and~Mugen~Peng,~\IEEEmembership{Fellow,~IEEE}}
\thanks{
This work was supported in part by the Beijing Municipal Science and Technology Project under Grant Z211100004421017, in part by the National Natural Science Foundation of China under Grants 62371071, and in part by the Young Elite Scientists Sponsorship Program by CAST under Grant 2021QNRC001. (\textbf{Corresponding author: Yaohua~Sun})

The authors are with the State Key Laboratory of Networking and Switching Technology, Beijing University of Posts and Telecommunications, Beijing 100876, China (e-mail:fengxinao@bupt.edu.cn; sunyaohua@bupt.edu.cn; pmg@bupt.edu.cn).}}
\maketitle
\begin{abstract}
Mega low earth orbit (LEO) satellite constellation is promising in achieving global coverage with high capacity. However, forwarding packets in mega constellation faces long end-to-end delay caused by multi-hop routing and high-complexity routing table construction, which will detrimentally impair the network transmission efficiency. To overcome this issue, a distributed low-complexity satellite-terrestrial cooperative routing approach is proposed in this paper, and its core idea is that each node forwards packets to next-hop node under the constraints of minimum end-to-end hop-count and queuing delay. Particularly, to achieve an accurate and low-complexity minimum end-to-end hop-count estimation in satellite-terrestrial cooperative routing scenario, we first introduce a satellite real-time position based graph (RTPG) to simplify the description of three-dimensional constellation, and further abstract RTPG into a key node based graph (KNBG). Considering the frequent regeneration of KNBG due to satellite movement, a low complexity generation method of KNBG is studied as well. Finally, utilizing KNBG as input, we design the minimum end-to-end hop-count estimation method (KNBG-MHCE). Meanwhile, the computational complexity, routing path survival probability and practical implementation of our proposal are all deeply discussed. Extensive simulations are also conducted in systems with Ka and laser band inter-satellite links to verify the superiority of our proposal.
\end{abstract}

\begin{IEEEkeywords}
LEO satellite network, mega constellation, minimum hop-count analysis, end-to-end routing.
\end{IEEEkeywords}

\section{Introduction}
\IEEEPARstart{W}{ith} the decreasing cost of satellite launches, mega  low earth orbit (LEO) satellite constellations have emerged as a viable solution to provide global broadband access\cite{sun2022,tao,TMC1,help1}. To further enhance communication capabilities, satellites will be equipped with on-board processing function and inter-satellite links (ISLs)\cite{LEOnetwork2021,help5}. In this case, a well-designed routing strategy is essential to fully utilize ISLs and guarantee the end-to-end quality of service\cite{help4,Timegraph2023,multipath,yuan2023}.

In general, existing routing strategies in LEO satellite constellations can be classified into centralized and distributed ones\cite{chen2022delay}. Centralized strategies typically leverage virtual topology\cite{DVTR1997} or virtual node\cite{VN1997} approaches to transform the satellite topology into a static one, and Dijkstra Shortest Path (DSP) algorithm can be executed on the static topology to obtain a routing path with minimum propagation delay.
Furthermore, as the traffic within constellation increases, routing strategy design should jointly considers propagation delay and queuing delay\cite{DISLOAD}. In this aspect, authors in \cite{Timegraph2023} propose a coordinate graph (CG) to characterize the satellite networks. By pruning and traversing the minimum-hop binary tree constructed by CG, a minimum end-to-end delay routing strategy is designed. Considering signal to noise ratio (SNR), link existence time, and the buffer queue size of ISLs, literature \cite{time2022} proposes a weighted time-space evolution graph to describe the topology of LEO satellite networks and designs a utility based routing algorithm. Although centralized routing strategy design can be simplified with snapshot idea\cite{graph1,graph2,graph3}, the complexity of them is high in mega constellations and updating routing tables for every satellite will also incur massive overhead\cite{Netgrid}.

Compared to centralized strategies, distributed routing strategies can make  more flexible and adaptive routing decisions based on collected global or local network information. When global network information is used for routing, each satellite maintains a real-time database which describes the status information of the whole network\cite{database}. To ensure the realtimeness of database, the state information of each satellite, such as available link ports and satellite load, need to be flooded through the network. However, due to the tens of milliseconds latency of ISLs, flooding among thousands of satellites will result in excessive routing convergence time. For example, the average route convergence time of Open Shortest Path First (OSPF) strategy in Iridium constellation is 38.43s\cite{flood1}, which will result in severe service interruption. To overcome this issue, literature \cite{GlobeDIS} stores the space-time graph on each satellite which can avoid the flooding of information on link changes caused by satellite movement. However, network-wide flooding is still required to address the unexpected link failures, and storing space-time graphs in all satellites will also lead to a significant consumption of storage resources.

On the other hand, routing based on local network information may not result in a globally optimal solution, but these strategies eliminate the need for network-wide flooding. Hence, fast routing convergence with acceptable overhead and low complexity can be achieved. In \cite{ELB2008}, authors propose Explicit Load Balancing (ELB) strategy, which forwards packets by considering the load status of next-hop satellite. However, it neglects the load status of sending buffer queue in current satellite, which may result in packet loss even the state of next-hop satellite shows "free". Moreover, Longer Side Priority (LSP) strategy \cite{LSP2019}  considers the load statuses of current satellite queues for packets forwarding, but the load status of next-hop satellite is not adequately taken into account. By jointly considering the load status of current satellite queues and next-hop satellite, Traffic-Light-Based routing strategy (TLR) \cite{TLR2014} realizes multi-path routing. However, the potential endless-loop problem in TLR will result in a large number of unnecessary routing hops in mega constellations.

In above routing strategies, they just utilize ISLs for routing, and applying those strategies in mega constellations will incur a significant increase in end-to-end routing hop-count. Based on the complex network theory, average end-to-end routing hops can be used to measure the network transmission efficiency, and bigger end-to-end routing hops always imply lower transmission efficiency and smaller network throughput \cite{res1,res2,res3}. Moreover, with the rapidly increasing traffic load in satellite network, too many routing hops will also result in long queuing delay \cite{res4,res5,res6}, which seriously affects the network transmission performance. Therefore, how to decrease the end-to-end routing hop-count becomes an important challenge \cite{res7,MinHop2021}. Most current research tries to reduce the end-to-end routing hops  and improve network performance by designing the constellation topology. Authors in literature \cite{res2} categorize satellites into backbone and access satellites, and dynamically adjust ISLs to enhance the network transmission efficiency. Similarly, literature \cite{res3} proposes a user-driven flexible and effective link connection method for mega constellations to reduce routing hops and improve network utilization. The authors in literature \cite{res9,res10} mention that temporary ISLs can be established between satellites moving in opposite directions, which can significantly reduce the average end-to-end routing hop-count of the network. All above studies establish dynamic and temporary ISLs to reduce end-to-end routing hops and improve network transmission performance. However, this is contrary to the dominant Iridium constellation and Starlink constellation \cite{res11,res12}, where each satellite will establish four permanent ISLs with its neighboring satellites and form a regular Manhattan network. In mega constellations, such Manhattan networks possess better stability, which can greatly reduce the topology dynamics and alleviate the network management overhead, as well as decrease the complexity of on-board transceivers design \cite{res9,res13,res14}. But the disadvantage is that this regular networking approach will cause some unnecessary routing hops and limit the network transmission performance. Therefore, by introducing a small number of ground relays, we plan to realize satellite-terrestrial cooperative routing, which can not only reduce the number of end-to-end routing hops, but also ensure the low dynamic advantage of regular Manhattan network.

In addition, to further ensure small end-to-end routing hops, each time forwarding can be achieved based on the minimum end-to-end hop-count constraint. Traditionally, minimum end-to-end hop-count constraint is got by network simulation\cite{MH2003}, \cite{MH2021}, but massive overhead and intolerable runtime will be a big problem in mega constellations. To solve this issue, an estimation method with $O(1)$ complexity is proposed in \cite{MinHop2021}. This method estimates the minimum end-to-end hop-count based on parameters of Walker constellation, but unfortunately it can not be extended to satellite-terrestrial cooperative routing scenario. Another feasible approach is executing Dijkstra on a two-dimensional (2D) graph built from three-dimensional (3D) constellation\cite{GraphMin}. However, if there are total $n$ satellites and $k$ ground relays, the complexity of one estimation is $O\big((n+k)^3\big)$, which lacks scalability in mega constellations.

Motivated by these observations, in this paper, we propose a low complexity distributed satellite-terrestrial cooperative routing approach, and its core idea is that each node forwards packets to next-hop node by jointly considering the minimum end-to-end hop-count and queuing delay constraints. Particularly, to achieve a low-complexity minimum end-to-end hop-count estimation in such cooperative routing scenario, satellite real-time position based graph (RTPG) is first designed to simplify the description of 3D constellation. By extracting key nodes in RTPG and generating edges with the minimum hop-count between nodes as weights, key node based graph (KNBG) is created. Next, considering the frequent regeneration of KNBG due to satellite movement, a low complexity generation method of KNBG is studied as well. Finally, with KNBG as input, a minimum end-to-end hop-count estimation method KNBG-MHCE is designed. Meanwhile, we fully discuss the property and practical implementation of our proposal, and extensive simulations are conducted to verify that our proposal can significantly reduce the end-to-end routing hop-count and enhance system throughput as well as delay performance.

The contributions of this paper are summarized as follows.
\begin{itemize}
\item \textbf{A Satellite-Terrestrial Cooperative Routing Framework and Low Complexity Minimum End-to-End Hop-Count Estimation Method: }We introduce a satellite-terrestrial cooperative routing framework, which consists of a mega LEO satellite constellation and a small number of ground relays. In this framework, each ground relay can establish satellite-ground links (SGLs) with multiple satellites within elevation range. By using SGLs, some pure inter-satellite forwarding processes can be replaced by satellite-terrestrial cooperative routing, which can significantly reduce the end-to-end routing hop-count under mega constellations. To further ensure small end-to-end routing hops, we design a  low-complexity method KNBG-MHCE to estimate the minimum end-to-end hop-count. Since the input of KNBG-MHCE is a key node based graph with lower dimensions, the complexity of KNBG-MHCE is reduced
    by a factor of 39.15 compared to traditional Dijkstra method in Starlink phase I constellation, which reflects the good scalability of our design.

\item \textbf{Low Overhead and Distributed Satellite-Terrestrial Cooperative Routing Strategy: }We propose a distributed satellite-terrestrial cooperative routing strategy based on the local load information and minimum end-to-end hop-count constraints given by KNBG-MHCE. To reduce the routing overhead, source routing paradigm is employed where the source node encapsulates the minimum end-to-end hop-count constraints derived from KNBG-MHCE into packets. Subsequently, each node can make distributed routing decisions with fast routing convergence by extracting the encapsulated information and jointly considering the load status of current satellite queues as well as next-hop node.

\item  \textbf{Deep Analysis of the Property, Practical Implementation, and Extensive Simulation of Our Proposal: }We analyze and compare the complexity of KNBG-MHCE with traditional Dijkstra method, and the path survival probability of our proposal is analyzed with inclusion-exclusion principle as well. Moreover, the conditions when the routing hop-count of the shortest distance path is equal to the minimum hop-count value is also discussed. To facilitate practical applications, we provide a detailed design of the packet encapsulation format to store the results of KNBG-MHCE. Finally, under Starlink phase I constellation, extensive simulations are conducted in systems with Ka and laser band ISLs to evaluate the performance of our proposal. Compared with routing strategies that just utilize ISLs, our proposal can reduce end-to-end delay by more than 200ms, and increase average system throughput by over 100\% in both systems. Compared with other satellite-terrestrial cooperative routing strategies, our proposal can improve system throughput by 13.29\% and 32.5\% under Ka and laser band ISLs, respectively.
\end{itemize}

The remainder of this paper is organized as follows. System model and satellite-terrestrial cooperative routing framework are described in Section \ref{sec:mod}. The graph model RTPG is introduced in Section \ref{sec:model1}. The entire KNBG-MHCE method is designed in Section \ref{sec:model2}. Our routing strategy is proposed in Section \ref{sec:model3}. Property analysis and packet format design are provided in Section \ref{sec:modanalysis}. Simulations are presented in Section \ref{sec:sim}, and Section \ref{sec:con} concludes the entire paper. The notation used in this paper is summarized in Table I.
\begin{table}[htpb]
\centering\caption{Notation}
\label{tab:notation}
\begin{tabular}{l l} \toprule 
\textbf{Notation} & \textbf{Definition} \\
\midrule 
$\omega_e$ & The angular velocity of the earth's rotation\\
$\omega_s$ & The angular velocity of satellite motion\\
$r_e $ & The radius of the earth\\
$N$ & The number of orbits in the constellation \\
$M$ & The number of satellites per orbit \\
$\Delta \Omega$ & The longitude difference between adjacent orbits \\
$\Delta \Phi$ & The phase difference between  neighboring satellites \\
& within a same orbit\\
$\Delta f$ & The phase difference between satellites connected by \\
& an inter-orbit link in adjacent orbits \\
$\xi(u_{m,t})$ & The longitude difference from orbit ascending node \\
& to satellite $m$ in the same orbit \\
$F$ &  The phasing factor of the constellation \\
$S$ & The total number of satellites in the constellation\\
$h$ & The orbit altitude \\
$\alpha$ & The orbit inclination with respect to the equator \\
$\lambda_{m,t},\varphi_{m,t}$ & The longitude and latitude of satellite $m$ \\
$u_{m,t}$ & The phase of satellite $m$ \\
$L_{n,t}$ & The longitude of ascending node in orbit $n$ \\
$H_{\text{min}}$ & Minimum end-to-end hops for inter-satellite routing\\
$R_{ISL}$ & The rate of inter-satellite link\\
$R_{SGL}^{up}, R_{SGL}^{down}$ & The rate of uplink and downlink satellite-ground link\\
$R_{pac}$ & The packet arrival rate\\
\bottomrule 
\end{tabular}
\end{table}
\section{System Model and Satellite-Terrestrial Cooperative Routing Framework}\label{sec:mod}
\subsection{Walker-Delta Constellation Model}
\begin{figure}[htbp]
\centering
\includegraphics[width = 0.9\linewidth]{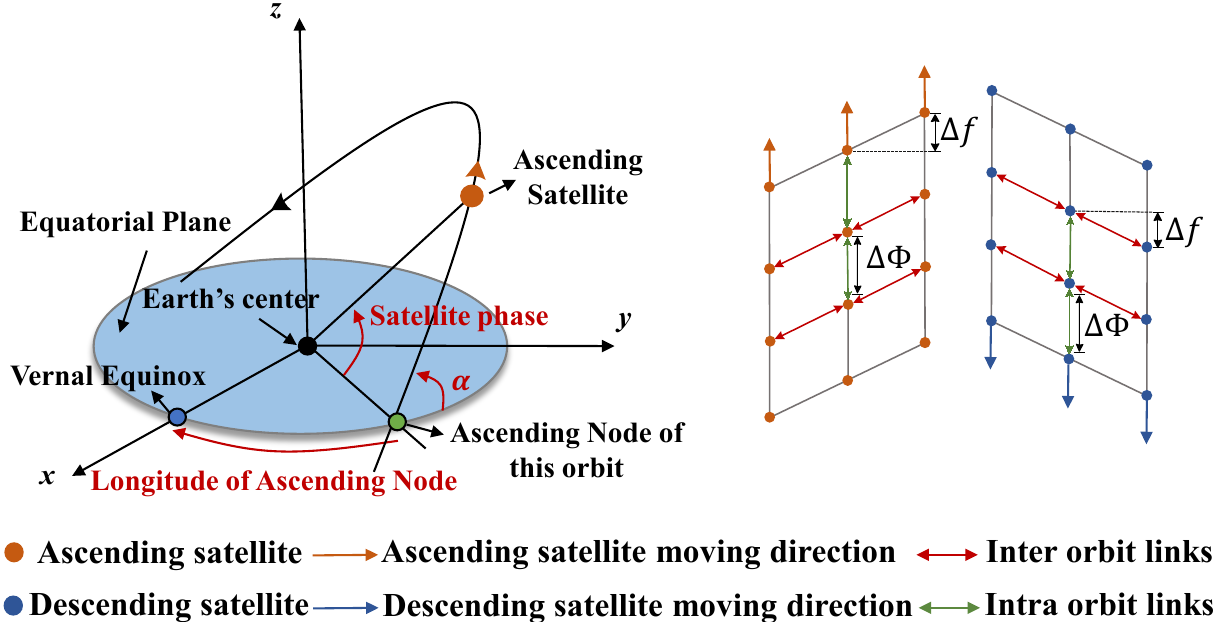}
\caption{Walker Delta constellation and ISL connection model.}
\label{figtop}
\end{figure}
{Our study primarily focuses on the mega Walker Delta constellation, which comprises $N$ orbits with an acute orbit inclination $\alpha$, and each orbit contains $M$ satellites.} As illustrated in Fig.~\ref{figtop}, $N$ orbits are uniformly distributed in the equatorial plane, and the difference between adjacent orbits in the longitude of ascending node is $\Delta \Omega = 2\pi / N$. Moreover, $M$ satellites are also uniformly distributed in each orbit, and the phase difference between neighboring satellites within an orbit is $\Delta \Phi = 2\pi / M$. Meanwhile, the phase difference between two satellites connected by an inter-orbit link is $\Delta f = 2\pi F / (NM)$, where $F$ is \textit{phasing factor} given as a constant in the design of constellation. In this paper, each satellite can establish four permanent ISLs with its neighboring satellites, consisting of two intra-orbit links and two inter-orbit links, and all satellites are further categorized as ascending satellites and descending satellites\cite{MinHop2021} based on their moving directions. Ascending satellites are moving in the latitude-increasing direction, while descending satellites are moving in the latitude-decreasing direction.
\subsection{ Satellite-Terrestrial Cooperative Routing Framework}\label{sec:mod2.2}
\begin{figure}[htbp]
\centering
\includegraphics[width = 0.95\linewidth]{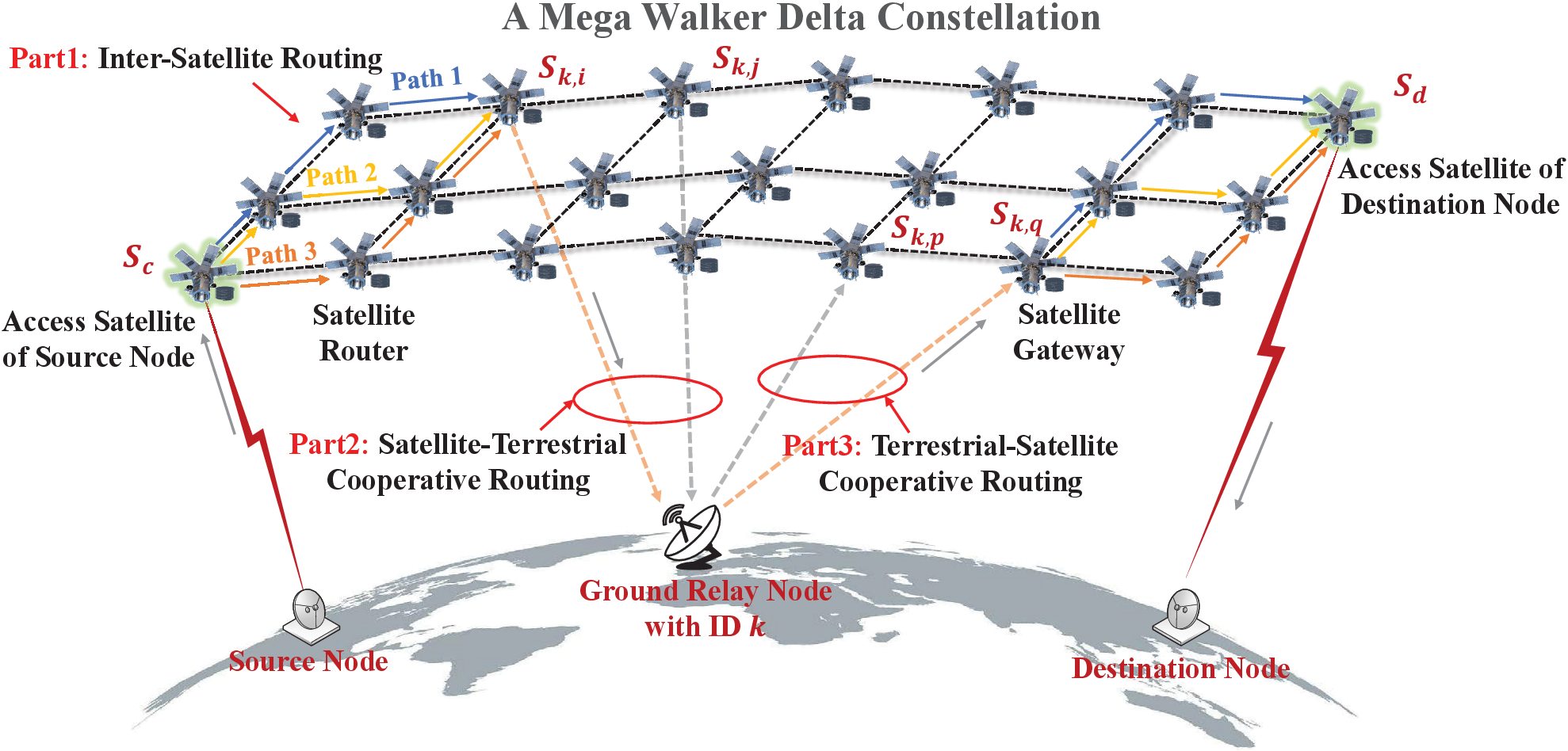}
\caption{Satellite-terrestrial cooperative routing framework.}
\label{figframe}
\end{figure}
The introduced satellite-terrestrial cooperative routing framework is shown in Fig.~\ref{figframe}, which consists of ground nodes and a mega Walker Delta constellation. Here, ground nodes include \emph{source nodes}, \emph{destination nodes}, and \emph{ground relay nodes}. {The source and destination nodes are accessed into the constellation as users\cite{Zhu}.} Every ground relay node can establish several SGLs with satellites within elevation range and make routing decisions according to routing strategy\cite{help2,yuan2024}. $GR_k$ is used to denote the ground relay whose ID is $k$. Moreover, all satellites in constellation possess on-board processing capability, enabling them to forward packets to neighboring nodes via ISLs or SGLs. In more detail, satellites that provide access service to source and destination nodes are defined as \emph{access satellite}, which are represented as $S_c$ and $S_d$, respectively. Satellites with SGLs are referred to \emph{satellite gateway}, and packets can be forwarded between satellite gateways and ground relays via SGLs. {For convenience, we denote the $i$-th satellite gateway belonging to ground relay $k$ as $SG_{k,i}$, and refer to all satellite gateways in system as \emph{key nodes}.} In addition, the selection of access satellites by source and destination nodes is outside the scope of our study, {so the end-to-end routing in the following means the routing process from $S_c$ to $S_d$.}

{Under the satellite-terrestrial cooperative routing framework, packets can be forwarded in three potential modes, which include \emph{inter-satellite forwarding}, \emph{satellite-terrestrial forwarding}, and \emph{terrestrial-satellite forwarding}. The last two modes represent the forwarding of packets between satellite gateway and ground relay via downlink SGL or uplink SGL, while the inter-satellite forwarding means packets are routed  between satellites via ISLs.} In this paper, each forwarding via ISL or SGL is defined as one hop, and the inter-satellite routing is performed based on the inter-satellite minimum hop-count estimation results, which specify the routing direction, as well as the minimum number of routing hops required in the horizontal and vertical direction. In Walker Delta constellation, all potential routing directions can be categorized into four types, which involve top-right, bottom-right, top-left, and bottom-left\cite{MinHop2021}. Taking Fig.~\ref{figframe} as an example, the minimum hop-count $H_{\text{min}}$ from $S_c$ to $S_{k,1}$ is 3, with a routing direction of top-right, where the minimum routing hops required in the horizontal and vertical direction are 1 and 2, respectively. Furthermore, in the inter-satellite routing mode with minimum hop-count constrains, each satellite can forward packets to the next-hop satellite in horizontal or vertical direction via ISLs, but when the remaining hop-count in one forwarding direction is 0, packets can only be forwarded to next-hop satellite corresponds to the other direction.

\begin{figure}[htbp]
\centering
\includegraphics[width = 0.8\linewidth]{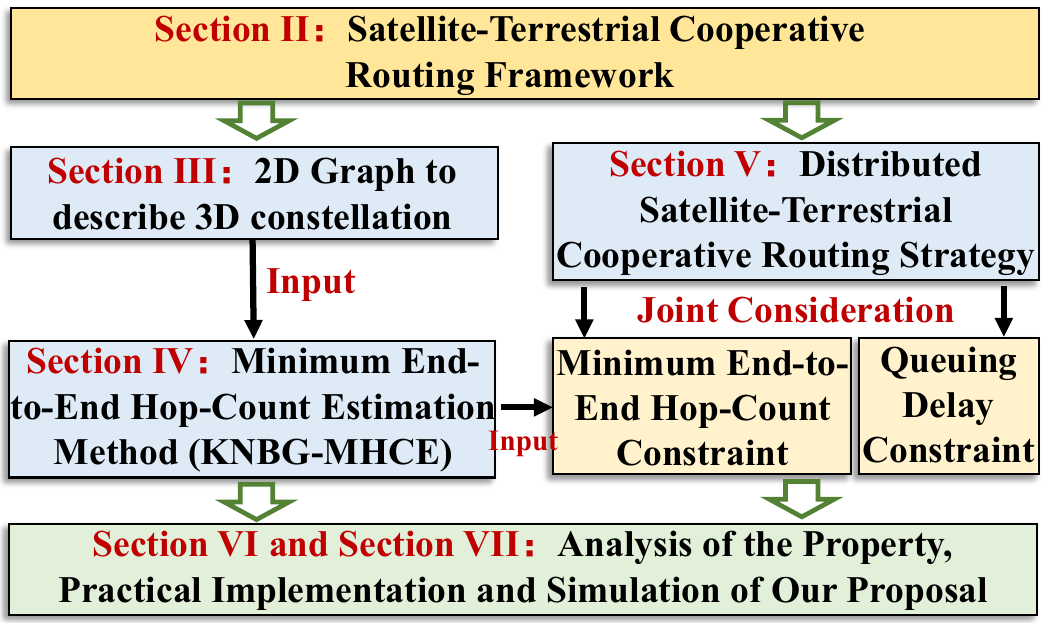}
\caption{The organization of entire paper.}
\label{figplan}
\end{figure}

Based on the framework described here, we deeply study the low-complexity minimum end-to-end hop-count estimation method as well as the satellite-terrestrial cooperative routing strategy, and the entire paper is organized as Fig.~\ref{figplan}. We first introduce the satellite real-time position based graph (RTPG) to describe 3D constellations in Section \ref{sec:model1}. By using RTPG, the minimum end-to-end hop-count estimation method (KNBG-MHCE) is designed in Section \ref{sec:model2}. Utilizing the minimum end-to-end hop-count constraints given by KNBG-MHCE and jointly considering the load status of satellites, we propose the distributed cooperative routing strategy in Section \ref{sec:model3}.
\section{{Two-Dimensional Graph Model Design}}\label{sec:model1}
\subsection{{Geographic Region Division of the Earth's Surface}}
In walker Delta constellation, since $N$ orbits are uniformly distributed along the equator at an interval of $\Delta \Omega$, {we divide the earth's surface into $N$ regions along the horizontal direction from $\lambda=0^\circ$}, and all regions have a width of $\Delta \Omega$. The longitudes corresponding to the left and right boundaries of region $P$ are given by \eqref{lambdalr}. {Moreover, satellites are uniformly distributed within their orbits with phase difference $\Delta \Phi$.} Therefore, each orbit can be divided into $M$ regions starting from the phase $\frac{\pi}{2}$. However, due to the non-linear relationship between the phase difference and latitude difference caused by orbit inclination, these regions will translate into $M$ unevenly distributed regions in vertical direction of the earth's surface. The latitudes corresponding to the lower and upper boundaries of region $R$ in vertical direction are given by \eqref{phidu}. {In the following two equations, $P$ takes value from $0, 1, \dots, N-1$, and $R$ takes values from $0, 1, \dots, M-1$.}
\begin{equation}
\begin{cases}
\lambda_{left} = P\cdot \Delta \Omega,\\
\lambda_{right} = (P+1)\cdot \Delta \Omega.
\end{cases}
\label{lambdalr}
\end{equation}
\begin{equation}
\begin{cases}
\varphi_{lower} = \text{arcsin}\big(\text{sin}\alpha\text{sin}\big( \frac{\pi}{2}-\frac{2\pi}{M}(M-R)\big) \big),\\
\varphi_{upper} = \text{arcsin}\big(\text{sin}\alpha\text{sin}\big( \frac{\pi}{2}-\frac{2\pi}{M}(M-R-1)\big) \big).
\end{cases}
\label{phidu}
\end{equation}

{By dividing the regions both horizontally and vertically,} the entire earth's surface is divided into $N \times M$ regions. Referring to Fig.~\ref{fig4}, when $N=5$ and $M=10$,  the earth's surface is divided into 5 regions in horizontal direction and 10 regions in vertical direction, {and each region can be represented by its coordinates $(P, R)$.} In addition, there is only one satellite within each region at any given time.
\subsection{Satellite Real-Time Position Based Graph Model}\label{sec:model12}
{The coordinates of region where each satellite is located at time $t$ need to be calculated, which is the key to convert the three-dimensional constellation into a two-dimensional graph.
 Although the position of satellites and orbits in relation to the earth is continuously changing, satellite phase and the longitude of ascending node can be used to indicate the position of a satellite, and determine which region the satellite is located in.} For the $m$-th satellite in orbit $n$ at time $t$, the latitude $\varphi_{m,t}$ and longitude $\lambda_{m,t}$ can be obtained according to the ephemeris information. {Employing these two parameters, the phase of satellite $m$ and the longitude of ascending node in orbit $n$ can be calculated by \eqref{equ1} and \eqref{equ2}, respectively\cite{MinHop2021}.}
\begin{equation}
{u_{m,t}} = \begin{cases}
\text{arcsin}\frac{\text{sin}\varphi_{m,t}}{\text{sin}\alpha},&{\text{ascending satellite,}} \\
\frac{\varphi_{m,t}}{|\varphi_{m,t}|}\pi - \text{arcsin}\frac{\text{sin}\varphi_{m,t}}{\text{sin}\alpha},&{\text{descending satellite.}}
\end{cases}
\label{equ1}
\end{equation}
\begin{equation}
    L_{n,t} = \mathbb{N}\big(\lambda_{m,t} - \xi(u_{m,t})\big).
\label{equ2}
\end{equation}
\begin{equation}
{\xi(u_{m,t})} =
\begin{cases}
\text{arctan}(\text{cos}\alpha \text{tan} u_{m,t}),&{\text{ascending satellite,}} \\
\text{arctan}(\text{cos}\alpha \text{tan}u_{m,t}) + \pi,&{\text{descending satellite.}}
\end{cases}
\label{equ3}
\end{equation}

$u_{m,t}\in [-\pi , \pi]$ represents the phase of satellite $m$ at time $t$. {When the phase of satellite is within $[-\pi/2 , \pi/2]$, it signifies that this satellite is an ascending satellite. Otherwise, this satellite is a descending satellite. On the other hand, $L_{n,t} \in [0 , 2\pi] $ represents the longitude of ascending node in orbit $n$. $\mathbb{N}(x) = \text{mod}(x,2\pi)$ in \eqref{equ2} is a normalisation function that ensures the value of $L_{n,t}$ is within the interval $[0 , 2\pi]$.
Based on $u_{m,t}$ and $L_{n,t}$, we can calculate coordinates $(P_{n,t}, R_{m,t})$ of the region where satellite $m$ in orbit $n$ is located at time $t$, and $P_{n,t}$ is expressed as
\begin{equation}
P_{n,t} = \big\lfloor \frac{L_{n,t}}{\Delta \Omega} \big\rfloor, \quad n = 1,2,\cdots,N.
\label{equPn}
\end{equation}

In equation \eqref{equPn}, $P_{n,t}$ takes values from $0, 1, \cdots, N-1$, and a {different value of $P_{n,t}$  indicates} that the satellite is located in different horizontal region. Since $L_{n,t} $ will be changed due to the earth rotation, $P_{n,t}$ is a periodic variable with period $T_e = 2\pi/\omega_e$, where $\omega_e$ is an angular velocity of the earth rotation. The other coordinate $R_{m,t}$ is given as
\begin{equation}
R_{m,t} = \big \lfloor \frac{\boldsymbol{u}_{m,t}- \pi/2}{\Delta \Phi} \big \rfloor, \quad m = 1,2,\cdots,M,
\label{equRm}
\end{equation}
\begin{equation}
\text{where }\boldsymbol{u}_{m,t} = \begin{cases}
u_{m,t},&{\text{if $u_{m,t} \geq \pi/2$,}} \\
u_{m,t} + 2\pi,&{\text{if $u_{m,t} < \pi/2$.}}
\end{cases}
\label{Ur}
\end{equation}

Following \eqref{equRm} and \eqref{Ur}, $R_{m,t}$ takes value from $0, 1, \cdots,M-1$, and a {different value of $R_{m,t}$ indicates} the satellite is located in different vertical region. Since all satellites move along the orbit with angular velocity $\omega_s$, $R_{m,t}$ is also a periodic variable with period $T_s = 2\pi / \omega_s$. Moreover, the phases corresponding to the lower and upper boundaries of region $R$ in vertical direction are expressed as
\begin{equation}
\begin{cases}
\mathcal{U}_{u}(\boldsymbol{u}_{m,t}) = (R_{m,t} + 1)\Delta \Phi,\\
\mathcal{U}_{l}(\boldsymbol{u}_{m,t}) =  R_{m,t}\Delta \Phi.
\label{R4}
\end{cases}
\end{equation}

By calculating coordinates of the region in which each satellite is positioned, we can transform the constellation into a two-dimensional graph model RTPG as shown in Fig.~\ref{fig4}.
\begin{figure}
\centering
\includegraphics[width = 0.8\linewidth]{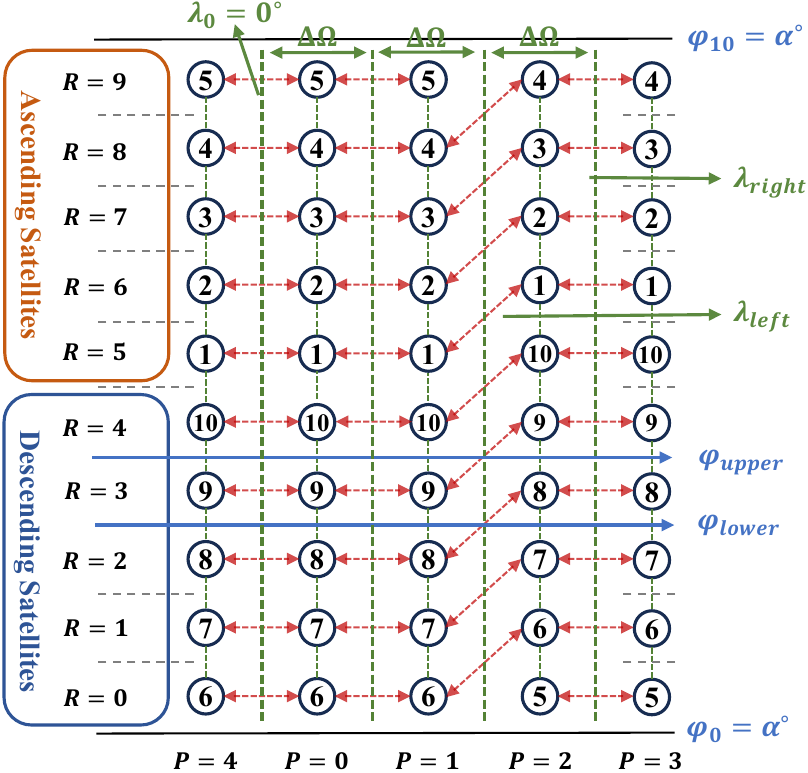}
\vspace{-0.1cm}
\caption{Walker Delta constellation topology based on RTPG model.}
\vspace{-0.3cm}
\label{fig4}
\centering
\end{figure}
\section{Minimum Hop-Count Estimation Method For Satellite-Terrestrial
Cooperative Routing}\label{sec:model2}
 Through the study of section \ref{sec:model1}, 3D constellation can be depicted as a 2D graph RTPG. In this section, by extracting key nodes in RTPG and generating weighted edges between them, we form the key node based graph (KNBG), and introduce KNBG based minimum end-to-end hop-count estimation method (KNBG-MHCE) for satellite-terrestrial cooperative routing. It is further note that KNBG will be regenerated frequently due to satellite movement, so we fully develop the key node extraction and weighted edge generation methods to realize low-complexity generation of KNBG.
\subsection{Minimum Hop-Count Estimation for Pure Inter-Satellite Forwarding Mode}\label{sec:model21}
{Before designing KNBG-MHCE method, we need to study the minimum end-to-end hop-count estimation between two satellites under inter-satellite forwarding mode at first. The estimation results will then serve as weights for the edges connecting key nodes within KNBG, and give the routing direction and minimum hop-count constraints for inter-satellite routing as well. Aiming at reducing the computational overhead, a minimum end-to-end hop-count estimation method with complexity $O(1)$ for inter-satellite forwarding mode is designed based on satellite coordinates in RTPG. For brevity, we assume the coordinates of two satellite nodes in RTPG are $(P_{n1,t}, R_{m1,t})$ and $(P_{n2,t}, R_{m2,t})$, and the minimum end-to-end hop-count between them in inter-satellite forwarding mode is $H_{\text{min}}$. Since the following derivations are applicable at any given time, variable $t$ is omitted.

As illustrated in Section \ref{sec:mod2.2}, there are four possible routing directions in inter-satellite forwarding mode. Hence, there will be a minimum end-to-end hop-count in each of these four directions, and $H_{\text{min}}$ is the smallest one of these four values, which can be expressed as follows\cite{MinHop2021}}
\begin{equation}
\begin{split}
H_{\text{min}} = \text{min} & \{H_h^{\rightarrow} + H_v^{\rightarrow \uparrow}, H_h^{\rightarrow} + H_v^{\rightarrow \downarrow}, \\
& H_h^{\leftarrow} + H_v^{\leftarrow \uparrow}, H_h^{\leftarrow} + H_v^{\leftarrow \downarrow}\}.
\label{Hmin}
\end{split}
\end{equation}
In equation \eqref{Hmin}, $H_h^{\rightarrow}$ and $H_h^{\leftarrow}$ denote the minimum forwarding hops required horizontally {to the right and left,} and {$H_v^{\rightarrow \uparrow}$ and $H_v^{\rightarrow \downarrow}$ indicate the minimum hop-count required upward and downward when the routing direction is rightward in horizontal direction. Similarly, $H_v^{\leftarrow \uparrow}$ and $H_v^{\leftarrow \downarrow}$ signify the minimum hop-count required upward and downward when the routing direction is leftward in the horizontal. With regards to obtain $H_{\text{min}}$, $H_h^{\rightarrow}$ and $H_h^{\leftarrow}$ are first calculated as follows}
\begin{subnumcases}
{}H_h^{\rightarrow} = \text{mod} \big( N - (P_{n1} - P_{n2}) , N\big), \label{Hh11}\\
H_h^{\leftarrow} = \text{mod} \big( N + (P_{n1} - P_{n2}) , N\big).
\label{Hh12}
\end{subnumcases}

{Second, \eqref{Hv2} and \eqref{Hv3} are functions describing the change in coordinate $R$ caused by the phase difference $\Delta f$ when forwarding horizontally to the right and left, respectively.
\begin{figure*}[!t]
\normalsize
\setcounter{MYtempeqncnt}{\value{equation}}
\setcounter{equation}{11}
\begin{equation}
\mathbb{R}(H_h^{\rightarrow} , R_{m1}) = \begin{cases}
R_{m1}, & H_h^{\rightarrow} \Delta f \leq \mathcal{U}_{upper}(\boldsymbol{u}_{m,t}), \\
\text{mod}\bigg(R_{m1} + \Big \lceil \frac{H_h^{\rightarrow} \Delta f - |\mathcal{U}_{upper}(\boldsymbol{u}_{m,t}) - \boldsymbol{u}_{m,t} |}{\Delta \Phi} \Big \rceil,  M\bigg), & H_h^{\rightarrow} \Delta f > \mathcal{U}_{upper}(\boldsymbol{u}_{m,t}).
\end{cases}
\label{Hv2}
\end{equation}
\setcounter{equation}{\value{MYtempeqncnt}}
\end{figure*}
\begin{figure*}[!t]
\normalsize
\setcounter{MYtempeqncnt}{\value{equation}}
\setcounter{equation}{12}
\begin{equation}
\mathbb{R}(H_h^{\leftarrow} , R_{m1}) = \begin{cases}
R_{m1}, & H_h^{\leftarrow} \Delta f \leq \mathcal{U}_{lower}(\boldsymbol{u}_{m,t}), \\
\text{mod}\bigg(R_{m1} + \Big \lceil \frac{H_h^{\leftarrow} \Delta f - |\mathcal{U}_{lower}(\boldsymbol{u}_{m,t}) - \boldsymbol{u}_{m,t} |}{\Delta \Phi} \Big \rceil, M\bigg), &  H_h^{\leftarrow} \Delta f > \mathcal{U}_{lower}(\boldsymbol{u}_{m,t}).
\end{cases}
\label{Hv3}
\end{equation}
\setcounter{equation}{\value{MYtempeqncnt}}
\hrulefill
\vspace*{1pt}
\end{figure*}

Substituting \eqref{Hv2} and \eqref{Hv3} into \eqref{Hv11} results in $H_v^{\rightarrow \uparrow}$ and $H_v^{\leftarrow \uparrow}$. In a similar way, substituting these two equations into \eqref{Hv12} yields $H_v^{\rightarrow \downarrow}$ and $H_v^{\leftarrow \downarrow}$.}
\setcounter{equation}{13}
\begin{subnumcases}
{}H_v^{\uparrow} = \text{mod} \big( M + \big(R_{m2} - \mathbb{R}(H_h, R_{m1})\big) , M\Big),\label{Hv11}\\
H_v^{\downarrow} = \text{mod} \big( M - \big(R_{m2} - \mathbb{R}(H_h, R_{m1})\big) , M\Big).
\label{Hv12}
\end{subnumcases}

Finally, based on derivations above, the minimum end-to-end hop-count $H_{\text{min}}$ and the corresponding routing direction for inter-satellite forwarding mode can be obtained according to \eqref{Hmin}, and the calculation process is summarized in Algorithm 1, which exhibits $O(1)$ complexity.
\begin{algorithm}[htbp]
\label{alg:1}
	\begin{algorithmic}[1]
		\caption{Minimum Hop-Count Estimation Method for Inter-Satellite Forwarding Mode}
        \Require {The ephemeris information dataset}
        \Ensure {$H_{min}$}
        \State Initialize $(P_{n1,t}, R_{m1,t})$ and $(P_{n2,t}, R_{m2,t})$ using ephemeris information;
        \State Calculate $H_h$ using \eqref{Hh11} and \eqref{Hh12};
        \State Evaluate $H_v$ using \eqref{Hv2}, \eqref{Hv3} and \eqref{Hv11}, \eqref{Hv12};
        \State Compute $H_{min}$ based on \eqref{Hmin}.
	\end{algorithmic}
\end{algorithm}
\vspace{-0.25cm}
\subsection{Minimum Hop-Count Estimation for Satellite-Terrestrial Cooperative Forwarding Mode}\label{sec:model22}
When one user accesses any satellite, this satellite is going to estimate the minimum end-to-end hop-count for these packets, and the estimation results will provide a minimum hop-count constraint for subsequent satellite-terrestrial cooperative routing. To realize this estimation, it is first necessary to periodically generate the key node based graph KNBG in every satellite, where key nodes refer to all satellite gateways.
The generation process of KNBG can be divided into two steps, which contain key node extraction and weighted edge generation. Due to satellite movement, satellite gateways in the system are not fixed, and thus KNBG face frequent regenerations. Therefore, we fully design above two steps to ensure the low complexity regeneration of KNBG. Next, taking KNBG as input, Dijkstra is executed to estimate the minimum end-to-end  hop-count. We refer to the method formed by above three steps as KNBG-MHCE, and a more comprehensive symbol description is provided in Section \ref{sec:mod2.2}.
\subsubsection{\textbf{{Key Nodes Extraction}}}
The first step in generating KNBG is to extract all key nodes in RTPG and get their coordinates $(P_{n}, R_{m})$. Since KNBG is regenerated frequently, how one satellite node obtains the coordinates of all key nodes with low complexity is the key to design. {In fact, satellite gateways are located within the elevation range of ground relays, and all satellite gateways belonging to $GR_k$ will be located within a circular area centered around $GR_k$ with $r_s$ as radius.} We define this circular area as \emph{search range}. {If the earth radius is $r_e$} and the minimum elevation angle of satellite is $\theta$, $r_s$ can be calculated as $r_s = r_e\cdot \beta$, where
\setcounter{equation}{14}
\begin{equation}
\beta = \frac{\pi}{2} - \theta - \gamma, \quad\gamma = \text{arcsin}\frac{r_e\cdot \text{sin}(\theta + \frac{\pi}{2})}{h+r_e}.
\end{equation}

{In traditional approach, to extract all key nodes in system, we need to calculate the elevation angle between ground relay $GR_k$ and all satellites. If the elevation angle between one satellite and $GR_k$ is bigger than the minimum elevation angle, this satellite becomes satellite gateway belonging to $GR_k$. When there are $S$ satellites and $K$ ground relays in the system, traditional approach requires a total of $KS$ computations. Since the coordinates of one satellite in RTPG can indicate which geographic region it is in, we can use this information to reduce the computational complexity of searching for key nodes. As shown in Fig.~\ref{fig5_1},} we can first equate $GR_k$ and its search range as node $GR_k^*$ and area $\mathcal{R}_k^*$ in RTPG, respectively. After that,
by searching for satellites within $\mathcal{R}_k^*$ {that satisfy the elevation requirements,} all satellite gateway $SG_{k, i}$ belonging to $GR_k$ can be obtained without the need to traverse entire constellation. It's worth noting that $GR_k^* $ has a coordinate $(P_{GR_k^* }, R_{GR_k^* })$ just like the satellite nodes in RTPG, and $\mathcal{R}_k^*$  is a rectangular area formed by several original regions in RTPG.

 We realize the above equivalence in two steps. First the coordinates of $GR_k^* $ are determined by \eqref{equPn} and \eqref{equRm}. It should be emphasized that according to \eqref{equ1}, there are two cases for the phase of $GR_k^*$, and therefore there will be two coordinates of $GR_k^*$. The reason for this phenomenon is that any ground relay node will be covered by both ascending and descending satellites. Consequently, $GR_k^*$ at different coordinates will search for satellite gateways in different moving direction. In the following, we denote the two coordinates of $GR_k^*$ as $(P_{GR_k^{A*} }, R_{GR_k^{A*} })$ and $(P_{GR_k^{D*} }, R_{GR_k^{D*} })$ respectively. Second, we derive the number of original regions occupied by $\mathcal{R}_k^*$ in the horizontal and vertical directions, which are denoted by $\Delta P$ and $\Delta R$. In the horizontal direction, since the widths of original regions in RTPG are all $\Delta \Omega$, $\Delta P$ is calculated as
 \begin{equation}
\begin{split}
    \Delta P = 2\cdot \big\lceil \frac{180\cdot r_s}{\pi \cdot r_e\text{cos}(\varphi_{GR_k})\Delta\Omega} \big\rceil.
\end{split}
\label{DeltaP}
\end{equation}

On the other hand, as described in Section \ref{sec:model12}, the original regions in RTPG have uneven heights in the vertical direction. {In order to obtain a concise expression for $\Delta R$, we assume that the heights of original region in RTPG are all $\Delta h_{min}$.} Therefore, the number of original regions occupied by $\mathcal{R}_k^*$ in vertical direction is given by
\begin{equation}
\Delta R = 2\cdot \big\lceil \frac{180\cdot r_s}{\pi \cdot r_e\cdot\Delta h_{min}}\big\rceil.
\label{deltar}
\end{equation}

{$\Delta h_{min}$ in \eqref{deltar} refers to the minimum value of the height of all regions in RTPG. To derive $\Delta h_{min}$, the height of region $R_m$ in RTPG is first defined as follows}
\begin{equation}
\Delta h = \big|f _{R_m\rightarrow\varphi}(R_m+1) - f _{R_m\rightarrow\varphi}(R_m)\big|.
\label{h}
\end{equation}
$f _{R_m\rightarrow\varphi}(R_m+1)$ and $f _{R_m\rightarrow\varphi}(R_m)$ in \eqref{h} represent the latitudes corresponding to the upper and lower boundaries of region $R_m$ in vertical direction, which can be written uniformly as
\begin{equation}
f _{R_m\rightarrow\varphi}(r)= \text{arcsin}\big(\text{sin}(\alpha)\text{sin}\big(\frac{\pi}{2} - \frac{2\pi}{M}\cdot(M-r)\big)\big),
\label{fR}
\end{equation}
 where $r=0, 1, \dots M$. By taking the derivative of \eqref{fR}, we can obtain
 \begin{equation}
\triangledown f _{R_m\rightarrow\varphi}(r)= \frac{\frac{2\pi}{M}\text{sin}(\alpha)\text{cos}(\frac{\pi}{2}-\frac{2\pi}{M}\cdot(M-r))}{\sqrt{1-\text{sin}^2(\alpha)\text{cos}^2(\frac{2\pi}{M}\cdot(M-r))}}.
\label{deltafR}
\end{equation}

{Since the minimum value of \eqref{deltafR} is taken at $r = 0$,} $\Delta h_{min}$ is also achieved at $r = 0$, which is calculated as
\begin{equation}
h_{min} = \alpha - \text{arcsin}\big(\text{sin}(\alpha)\text{sin}\big(\frac{\pi}{2} - \frac{2\pi}{M}\cdot(M-1)\big)\big).
\label{hmin}
\end{equation}

{Substituting \eqref{hmin} into \eqref{deltar}, and utilizing \eqref{DeltaP} and \eqref{deltar} to calculate $\Delta P$ and $\Delta R$, we can obtain the coordinates of original regions in RTPG covered by $\mathcal{R}_k^*$.} The range of horizontal and vertical coordinates of these regions are expressed as \eqref{hsearch} and \eqref{vsearch}, respectively. By searching regions within \eqref{hsearch} and \eqref{vsearch}, all key nodes can be extracted, and the number of computations will be much smaller than the traditional approach.
\begin{figure*}[!t]
\normalsize
\setcounter{MYtempeqncnt}{\value{equation}}
\setcounter{equation}{21}
\begin{equation}
\begin{cases}
\big[\text{mod}\big(P_{GR_k^{A*}} - \frac{\Delta P}{2} + N, N\big), \text{mod}\big(P_{GR_k^{A*}} + \frac{\Delta P}{2}, N\big)\big],&{\text{search for ascending satellite gateways,}} \\
\big[\text{mod}\big(P_{GR_k^{D*}} - \frac{\Delta P}{2} + N, N\big), \text{mod}\big(P_{GR_k^{D*}} + \frac{\Delta P}{2}, N\big)\big],&{\text{search for descending satellite gateways.}} \\
\end{cases}
\label{hsearch}
\end{equation}
\setcounter{equation}{\value{MYtempeqncnt}}
\end{figure*}
\begin{figure*}[!t]
\normalsize
\setcounter{MYtempeqncnt}{\value{equation}}
\setcounter{equation}{22}
\begin{equation}
\Big[\text{mod}\big(R_{GR_k^*} - \frac{\Delta R}{2} + M, M\big), \text{mod}\big(R_{GR_k^*} + \frac{\Delta R}{2}, M\big)\Big], \text{search for ascending and descending satellite gateways.}
\label{vsearch}
\end{equation}
\setcounter{equation}{\value{MYtempeqncnt}}
\hrulefill
\vspace*{1pt}
\end{figure*}
 \begin{figure}
\centering
\includegraphics[width = 0.9\linewidth]{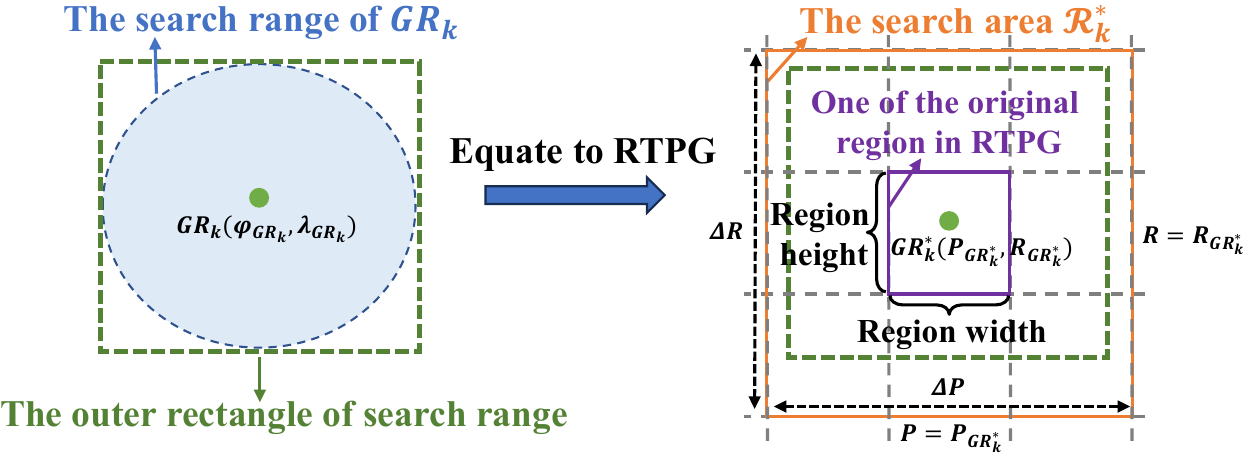}
\caption{Search area $\mathcal{R}_k^*$ and equivalent node $GR_k^* $ in RTPG model.}
\vspace{-0.5cm}
\label{fig5_1}
\centering
\end{figure}
\subsubsection{\textbf{Weighted Edge Generation}}
{For practical implementation, we define a unique ID for each satellite. Satellite $m$ in orbit $n$ will be denoted as $n_i$,} where $i$ is the satellite ID calculated by $i=(n-1)\times M + m$. Moreover, the weight of an edge between a pair of key nodes is defined as $w_{i,j}$, which is the minimum hop-count between them and can be calculated as follows.
\begin{itemize}
\item \emph{$w_{i,j}=2$}: $n_i$ and $n_j$ are satellite gateways belonging to the same ground relay $GR_k$
\item \emph{$w_{i,j}=H_{min}$}: $n_i$ and $n_j$ are satellite gateways belonging to the same ground relay $GR_k$, but $H_{\text{min}}$ calculated by Algorithm 1 is smaller than 2. Or $n_i$ and $n_j$ are not satellite gateways belonging to the same $GR_k$.
\end{itemize}
\subsubsection{\textbf{Dijkstra with KNBG as Input }}
When one satellite provides access service to users, it becomes source satellite $S_c$. Next, nodes $n_c$, $n_d$ corresponding to $S_c$, $S_d$ are added to KNBG and the weighted edges between $n_c$, $n_d$ and other key nodes within KNBG are generated following the aforementioned weighted edge generation method. Finally, the minimum hop-count can be estimated by running Dijkstra with current KNBG as input, and the entire KNBG-MHCE method is demonstrated in Algorithm 2.
\begin{algorithm}[htbp]
	\begin{algorithmic}[1]
		\caption{KNBG-MHCE Method}
		\label{alg:1}
        \Ensure {all IDs of key nodes on the end-to-end routing path, and the path information between neighboring nodes.}
        \If {\text{mod} $(t,T) = 0$}
        \Comment{Executed on every satellite}
        \State  Initialize $KN = [\;]$ to store the IDs of all key nodes.
        \For {$k = 1:K$}
        \For {$Satellite(p,r)$ in \eqref{hsearch} and \eqref{vsearch}}
        \If {the elevation angle between $Satellite(p,r)$ and $GR_k$ meets the elevation constraint}
        \State $KN=[KN;\text{the ID of } satellite(p,r)]$
        \EndIf
        \EndFor
        \EndFor
        \State Initialize $w= [\;]$ to store the weights $w_{ij}$
        \For {$i = 1:\text{length}(KN)$}
        \For {$j = 1:\text{length}(KN)$}
        \State Calculate $w_{i,j}$, and store it in $w = [w;w_{i,j}]$
        \EndFor
        \EndFor
        \State Generate KNBG every $T$ period based on $KN$ and $w$
        \EndIf
        \If {one user accesses this satellite and its destination is $S_d$}
        \State $KN^*= [\text{the ID of } S_c;\text{the ID of } S_d]$
        \State Initialize $w^*= [\;]$ to store the weights $w^*_{ij}$
        \For {$i = 1:\text{length}(KN)$}
        \For {$j = 1:\text{length}(KN^*)$}
        \State calculate $w^*_{i,j}$, and store it in $w^* = [w^*;w^*_{i,j}]$
        \EndFor
        \EndFor
        \State Add $KN^*$ and $w^*$ to update KNBG, then run Dijkstra with KNBG as input
        \EndIf
	\end{algorithmic}
\end{algorithm}
\section{Minimum Hop-Count Based Routing Strategy for Satellite-Terrestrial Cooperative Routing}\label{sec:model3}
The results of KNBG-MHCE include all IDs of key nodes on the end-to-end routing path and the path information between neighboring nodes. In fact, the end-to-end routing path is divided into multiple segments by key nodes, if the path between two neighboring nodes is an inter-satellite routing path, its path information should contain routing direction, minimum routing hops in both horizontal and vertical directions. If the path between two neighboring nodes is a satellite-terrestrial cooperative routing path, the path information should contain the ID of utilized ground relay. By using these estimation results given by  KNBG-MHCE, we propose a distributed satellite-terrestrial cooperative routing strategy and its core idea is that each node forwards packets to next-hop node under the constraints of minimum end-to-end hop-count and queuing delay. For brevity, we refer to the proposed routing strategy as ``our proposal", or simply ``ours" in the following paper.
For all ISLs and downlink SGLs, a sending buffer queue $B_s$ is allocated to temporarily store packets that need to be forwarded to the next hop. Since ground relay nodes have sufficient storage space, {the sending buffer queue size for uplink SGL from ground relay to satellite is unlimited.} Additionally, a public waiting buffer queue $B_w$ is introduced according to \cite{TLR2014}, and link rates for ISL, uplink SGL, downlink SGL are denoted as $R_{ISL}$, $R^{up}_{SGL}$ and $R^{down}_{SGL}$, respectively.
\subsection{Our Satellite-Terrestrial Cooperative Routing Strategy}\label{sec:ours}

\subsubsection{\textbf{Inter-Satellite Forwarding}}In inter-satellite forwarding mode, packets are routed between satellites via ISLs. Referring to Section \ref{sec:mod2.2}, some terms are further defined as follows.
\begin{itemize}
\item \emph{candidate forwarding directions}: {Each satellite has two candidate forwarding directions, which are determined by the routing direction described in Section \ref{sec:mod2.2}. These two candidate forwarding directions consist of a horizontal direction and a vertical direction, and all optional combinations of them are: right \& up, right \& down, left \& up, and left \& down. Further note that the two candidate forwarding directions for one traffic flow are the same in all satellites on one inter-satellite routing path.}
\item \emph{candidate sending buffer queues}: {Two sending buffer queues} in satellite nodes corresponding to the two candidate forwarding directions.
\item \emph{next-hop satellite}: {Two neighboring satellites connected to current satellite via ISL in the candidate forwarding directions.}
\end{itemize}

When making routing decisions, the load status of current satellite queues and next-hop satellite are both considered. Assume that the load status of candidate sending buffer queue in current satellite $i$  in horizontal and vertical directions are $Q_i^h$ and $Q_i^v$, respectively. The load status of corresponding next-hop satellites in horizontal and vertical directions are $Q_{i+1}^h$ and $Q_{i+1}^v$. The load status of candidate sending buffer queue in next-hop satellite are $Q_{i+1,h}^h,Q_{i+1,v}^h$, and $Q_{i+1,h}^v,Q_{i+1,v}^v$. Then, $Q_{i+1}^h$ and $Q_{i+1}^v$ can be quantified as $Q_{i+1}^h = \big(Q_{i+1,h}^h + Q_{i+1,v}^h\big)/2$, $Q_{i+1}^v = \big(Q_{i+1,h}^v + Q_{i+1,v}^v\big)/2$.

By jointly considering the load status of candidate sending buffer queues in current satellite $i$ and the load
status of  next-hop satellites, delay cost metrics $T_i^h$ and $T_i^v$ are defined, which represent the potential queuing delay for forwarding in the horizontal and vertical directions.
\setcounter{equation}{23}
\begin{equation}
\begin{cases}
T_{i}^h = \big( Q_{i,h} + Q_{i+1}^{h}\big)/ R_{ISL}, \\
T_{i}^v = \big( Q_{i,v} + Q_{i+1}^{v}\big)/ R_{ISL}.
\label{Qi+1}
\end{cases}
\end{equation}

In our proposal, once queue saturation occurs in one candidate direction, next-hop satellite corresponding to the other candidate direction will be selected as the next-hop node. For example, {if $Q_i^h \geq B_s$ or $Q_{i+1}^h \geq B_s$, and both $Q_i^v$ and $Q_{i+1}^2$ are less than $B_s$, we call the sending buffer queue is saturated in horizontal candidate forwarding direction.} Then the satellite corresponding to vertical direction will be selected as the next-hop satellite. {Otherwise, routing decisions are made by comparing $T_i^h$ and $T_i^v$ with threshold $\Gamma$, which can be expressed as \eqref{gamma} and $\eta_1$ is a smoothing factor. This threshold is designed based on the remaining queue space for reducing the generation of long queuing delay caused by insufficient candidate directions. Insufficient candidate directions means that when the number of remaining hops in one candidate forwarding direction ($H^r_h$ or $H^r_v$) is 0, this packet can only be forwarded in the other direction. If there is severe congestion in that direction, it will lead to significant queuing delay. To address this issue, the number of remaining hop-count in candidate forwarding directions are also considered when comparing $T_i^h$ and $T_i^v$. For instance, if the remaining hop-count in horizontal candidate direction is less than that in vertical direction, we forward this packet in vertical direction as long as $T_i^h-T_i^v \leq \Gamma$. }
\begin{equation}
\Gamma = \eta_1 \times \big( \frac{2B_s}{R_{ISL}} - \text{max}\big( T_i^h, T_i^v\big)\big).
\label{gamma}
\end{equation}
\subsubsection{\textbf{Satellite-Terrestrial Forwarding}}Taking Fig.~\ref{figframe} as an example, when a packet has been forwarded to satellite gateway $SG_{k,i}$, if the load status of $SG_{k,i}$'s downlink SGL does not exceed threshold $\Psi_{down}$, ground relay $GR_k$ will be selected as the next-hop node, which is determined by KNBG-MHCE method. Alternatively, the packet will be first forwarded to another satellite gateway $SG_{k,j}$,  and then forwarded from $SG_{k,j}$ to $GR_k$. {$SG_{k,j}$ is a satellite gateway which has the lightest  load status on its downlink SGL among all satellite gateways belonging to $GR_k$, and the coordinates of $SG_{k,j}$ and $SG_{k,i}$ satisfy $| P_{SG_{k,i}} - P_{SG_{k,j}}| \leq N_0$.} $N_0$ is a constant set according to constellation size, which ensures that forwarding from $SG_{k,i}$ to $SG_{k,j}$ will not generate too many extra routing hops. {In contrast to forwarding packets from $SG_{k,i}$ to $GR_k$, forwarding from $SG_{k,j}$ to $GR_k$ will reduces the queuing delay on downlink SGL, but a few number of inter-satellite forwarding hops will be introduced due to the routing from $SG_{k,i}$ to $SG_{k,j}$.
Therefore, the design of threshold $\Psi_{down}$ requires a joint consideration of the transmission rate of ISL and downlink SGL,} which can be represented as follows, where $\eta_2$ and $\eta_3$ are smoothing factors and $\tau$ is time interval.
\begin{equation}
\Psi_{down} = \eta_2\times \tau R_{ISL} + \eta_3\times \tau R_{SGL}^{down}.
\end{equation}
\subsubsection{\textbf{Terrestrial-Satellite Forwarding}}Still using Fig.~\ref{figframe} as an example, when a packet has been forwarded to ground relay $GR_k$, if the load status of uplink SGL between $GR_k$ and $SG_{k,p}$ does not exceed threshold $\Psi_{up}$, this packet will be forward to satellite gateway $SG_{k,p}$, which is determined by KNBG-MHCE method. Otherwise, $GR_k$ will forward this packet to another satellite gateway $SG_{k,q}$. {$SG_{k,q}$ is a satellite gateway which has the lightest load status on the uplink SGL among all satellite
gateways belonging to $GR_k$, and the coordinates of $SG_{k,q}$ and $SG_{k,p}$ should also satisfy $| P_{SG_{k,q}} - P_{SG_{k,p}}| \leq N_0$. Since the routing strategy in terrestrial-satellite forwarding mode is similarly to that in satellite-terrestrial forwarding mode, the threshold $\Psi_{up}$ is designed  with reference to $\Psi_{down}$, which can be expressed as follows, and $\eta_4$ and $\eta_5$ are both smoothing factors.}
\begin{equation}
\Psi_{up} =
\eta_4\times \tau R_{ISL} + \eta_5\times \tau R_{SGL}^{up}.
\label{Psiup}
\end{equation}
\begin{algorithm}[htbp]
	\begin{algorithmic}[1]
		\caption{Satellite-Terrestrial Cooperative Routing Strategy}
		\label{alg:3}
        \Ensure {The ID of next-hop node}
        \If {packet is in the inter-satellite forwarding mode}
        \If {$H_v^r = 0$, $H_h^r \neq 0$ \textbf{or} $H_v^r \neq 0$, $H_h^r = 0$}
        \State next-hop is the satellite corresponding to candidate direction with $H^r \neq 0$.
        \ElsIf {horizontal queue saturated \textbf{or} vertical queue saturated \textbf{and} the other queue is unsaturated}
        \State next-hop is the satellite corresponding to the candidate forwarding direction with unsaturated queue.
        \Else
        \If {$H_h^r \leq H_v^r$, $T_i^h - T_i^v \leq \Gamma$ \textbf{or} $H_v^r \leq H_h^r$, $T_i^v - T_i^h > \Gamma$}
        \State next-hop is the satellite corresponding to the vertical candidate direction.
        \ElsIf {$H_h^r \leq H_v^r$, $T_i^h - T_i^v > \Gamma$ \textbf{or} $H_v^r \leq H_h^r$, $T_i^v - T_i^h \leq \Gamma$}
        \State next-hop is the satellite corresponding to the horizontal candidate direction.
        \EndIf
        \EndIf
        \EndIf
        \If {packet is in the satellite-terrestrial forwarding mode}
        \If { the queuing delay of $SG_{k,i}$'s downlink SGL does not exceed $\Psi_{down}$}
        \State $GR_k$ is selected as the next-hop node.
        \Else
        \State choose another satellite gateway $SG_{k,j}$, execute Algorithm 1 with coordinates of $SG_{k,i}$ and $SG_{k,j}$ as input, and then get the next-hop node according to the strategy of inter-satellite forwarding mode.
        \EndIf
        \EndIf
        \If {packet is in the terrestrial-satellite forwarding mode}
        \If {the queuing delay of uplink SGL between $GR_k$ and $SG_{k,p}$ does not exceed $\Psi_{up}$}
        \State $SG_{k,p}$ is selected as the next-hop node.
        \Else
        \State choose another satellite gateway $SG_{k,q}$ as the the next-hop node.
        \EndIf
        \EndIf
        \State \Return{the ID of next-hop node}
	\end{algorithmic}
\end{algorithm}
\section{{Property Analysis and Practical Implementation of Our Proposal}}\label{sec:modanalysis}
\subsection{{Analysis of Computational Complexity}}
In Section \ref{sec:model21}, KNBG-MHCE method (Algorithm 2) is designed to estimate the minimum end-to-end hop-count for satellite-terrestrial cooperative routing. This method consists of key node extraction, weighted edge generation, and execution of Dijkstra algorithm. Specifically, the complexity of key node extraction is $\mathcal{O}(2K\Delta P\Delta R)$, where $\Delta P$ and $\Delta R$ are calculated by \eqref{DeltaP} and \eqref{deltar}, and $K$ is the number of ground relay nodes. The complexity of weighted edge generation is $\mathcal{O}(K \frac{\mathcal{X}!}{2(\mathcal{X}-2)!}+2K\mathcal{X})$, where $\mathcal{X}$ is the average number of satellite gateways owned by each ground relay. The above two steps can form key node based graph KNBG, which contains $S^*$ nodes, and $S^*=K\mathcal{X}+2$ is the total number of nodes in KNBG. Running Dijkstra with KNBG as input can estimate the minimum end-to-end hop-count, and the complexity of this step is $\mathcal{O}\big((S^* )^2 \big)$. Therefore, the overall complexity of KNBG-MHCE is
$\mathcal{O}\big((K+ 0.5)K\mathcal{X}^2 + 5.5K\mathcal{X} +2K\Delta P\Delta R + 4\big) \cong \mathcal{O}\big((K+ 0.5)K\mathcal{X}^2 + 5.5K\mathcal{X} \big)\cong \mathcal{O}\big(f_1(K\mathcal{X})\big)$.

When using traditional Dijkstra method for minimum hop-count estimation, the first step is searching all satellite gateways in the network, and its time complexity is $\mathcal{O}(KS)$. Next, weighted edges between all nodes need to be created, and this step will generate graph $G$, which contains $S$ nodes and $S$ is the satellite number in system. The time complexity of this step is $\mathcal{O}\big(K\frac{\mathcal{X}!}{2(\mathcal{X}-2)!}\big)$. Next, running Dijkstra with $G$ as input, the minimum end-to-end hop-count estimation can be achieved, and the time complexity of this step is $O(S^2)$. Finally, the overall complexity of this method is $\mathcal{O}\big(KS + K\frac{\mathcal{X}!}{2(\mathcal{X}-2)!} + S^2\big) \cong \mathcal{O}\big(f_2(K\mathcal{X})\big)$.

Moreover, we define $\rho$ to indicate the proportion by which $f_1(K\mathcal{X})$ decreases compared to $f_2(K\mathcal{X})$, and it is given by
\begin{equation}
     \rho = \frac{f_{2}(K\mathcal{X})}{f_{1}(K\mathcal{X})} =
 \frac{KS + K\frac{\mathcal{X}!}{2(\mathcal{X}-2)!} + S^2}{(K+ 0.5)K\mathcal{X}^2 + 5.5K\mathcal{X}}.
    \label{rho}
\end{equation}

In the Starlink Phase I constellation of 1,584 satellites, we set 25 ground relays with average latitude of $28^\circ$, and $\mathcal{X}=10$ is obtained based on simulations. In this case, $\rho = 39.15$ is calculated according to \eqref{rho}, which means that estimating the minimum end-to-end hop-count with proposed KNBG-MHCE method can reduce the computational complexity by one order of magnitude compared to the traditional approach.
\subsection{{Analysis of Path Survival Probability}}
 Considering packets may be dropped due to congestion or link unavailability, path survival probability is introduced to quantify the likelihood of a packet successfully reaching its intended destination. In the following, we assume the link availability of inter-orbit links, intra-orbit links, and SGLs as $p$, $q$, $r$ respectively, and the path survival probability under different routing strategies are all analyzed in detail.
\subsubsection{\textbf{Path Survival Probability of Dijkstra Shortest Path (DSP) and DisCoRoute (DIS)}}
DSP and DIS\cite{DIS2022} will establish a single routing path. and the end-to-end path survival probability can be written as $P_{1} = p^{H_h}q^{H_v}$.
\subsubsection{\textbf{Path Survival Probability of Longer Side Priority (LSP)}}
When the number of remaining hops in the horizontal and vertical forwarding directions are not equal, LSP\cite{LSP2019} forwards packets to the direction with more remaining hops. When these two values are equal, LSP forwards packets to the direction with lighter queue load, so the path survival probability under LSP strategy can be calculated as
\begin{equation}
P_2 =
\begin{cases}
p^{H_h-H_v}\big(pq(2-pq)\big)^{H_v}, &{\text{if $H_h \geq H_v$,}}\\
q^{H_v-H_h}\big(pq(2-pq)\big)^{H_h}, &{\text{if $H_h < H_v$.}}
\label{P2}
\end{cases}
\end{equation}
\subsubsection{\textbf{Path Survival Probability of Our Proposal}}
In inter-satellite forwarding mode, our proposal enables multi-path routing based on $H_{\text{min}}$, and the number of routing paths that satisfy the minimum end-to-end hop-count constrains can be calculated by
\begin{equation}
\centering
N_p = \frac{H_{\text{min}}!}{H_v!\cdot (H_{\text{min}}-H_v)!}.
\end{equation}
Based on the the inclusion-exclusion principle, path survival probability in this mode can be calculated by \eqref{pathInter}, where $A_{a_i}$ denotes the $A_{a_i}$-th routing path.
\begin{equation}
    P_{InterSat} = \sum_{m=1}^{N_P} (-1)^{m-1}\sum_{1 \leq a_i < a_{i+1} \leq N_p}\Big| P\Big(\bigcap_{a_1}^{a_m} A_{a_i}\Big) \Big|.
    \label{pathInter}
\end{equation}
{In cooperative forwarding mode,} $H_g$ is used to represent the number of satellite-ground hops. As a result, the path survival probability under our proposal is $P_{3} = P_{InterSat} \cdot r^{H_g}$.
\begin{figure}[htpb]
\centering
\includegraphics[width=0.45\textwidth]{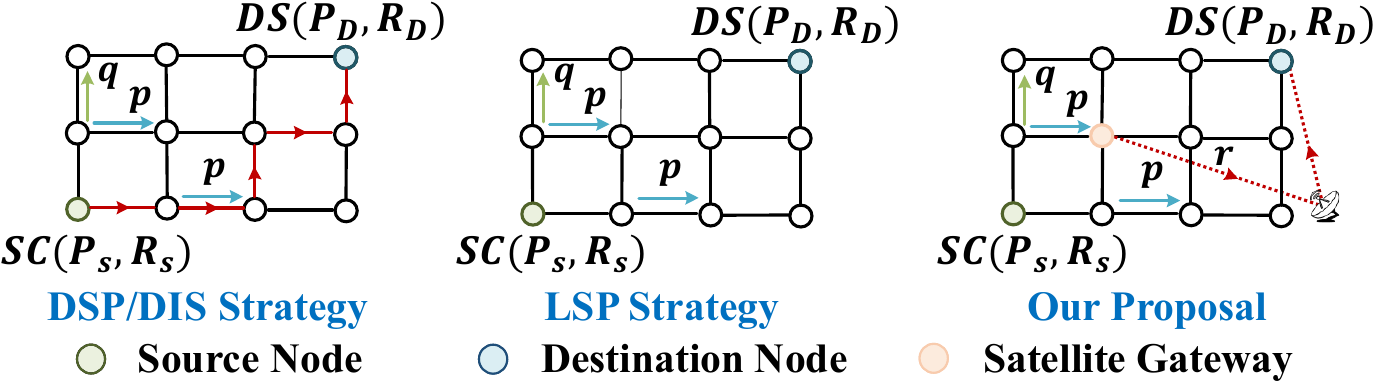}
\caption{The routing paths formed by different routing strategies.}\label{fig12}
\end{figure}

Taking Fig.~\ref{fig12} as an example, $P_1, P_2$ and $P_3$ are
\begin{equation}
P_{1} = p^3q^2, ~ P_{2} = p^3q^2(2-pq)^2, ~ P_{3} = pq(2-pq)r^2.\\
\label{Ppath}
\end{equation}

Based on \eqref{Ppath}, we conduct simulations to assess $P_1, P_2$ and $P_3$ under various values of $p$ and $q$. Since only $P_3$ is related to $r$, $r$ is set to constant 1, and we only compare the effect caused by the changes in link available probability of ISLs. Meanwhile, the simulation results are shown in Fig.~\ref{figPSP}. As $p$ and $q$ increase, the link availability is enhanced, resulting in an upward trend of path survival probability for all strategies. In all cases, our proposal exhibits the highest path survival probability attributed to the minimum hop-count and its congestion avoidance mechanism.
\begin{figure}[htpb]
\centering
\includegraphics[width=0.45\textwidth]{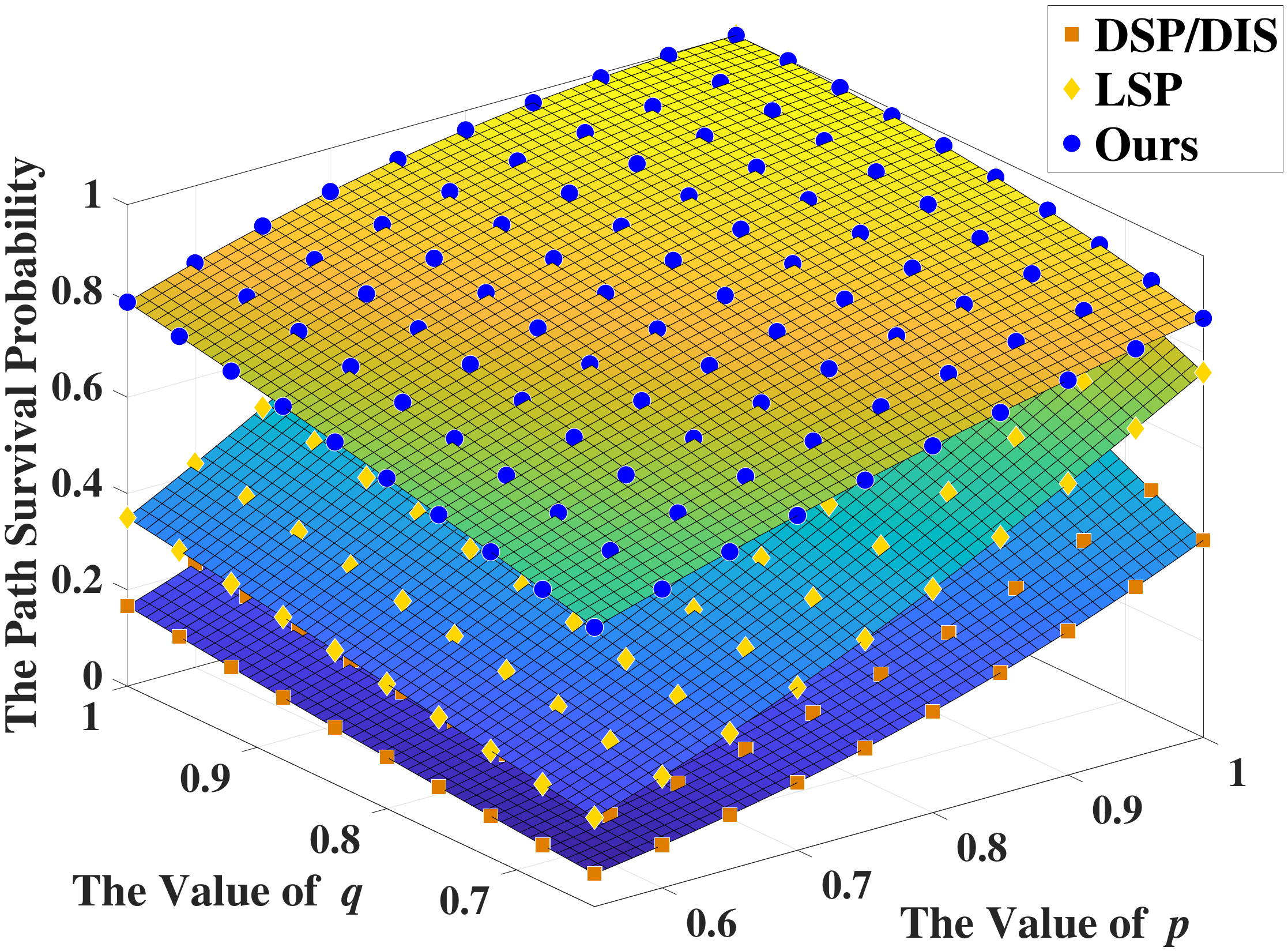}
\caption{The path survival probability of different strategies.}
\label{figPSP}
\end{figure}
\subsection{{Analysis of Condition When Routing Hop-Count of the Shortest Distance Path is Equal to the Minimum Value}}\label{hopcount}
Minimum hop-count and shortest distance path are two commonly used concepts in routing strategy design. Generally, there may be no direct correlation between these two routing paths. However, in mega Walker Delta constellations, the end to-end routing hops of the shortest distance path tends to be the minimum value. Assuming that the Walker Delta constellation comprises $N$ orbits, while each orbit contains $M$ satellites and the phasing factor of this constellation is $F$. $P_s$ denotes the shortest distance path and $P_{mh}$ denotes the minimum hop-count path. To obtain the sufficient conditions for the end-to-end routing hops of the shortest distance path $H_{s}$ equals to the minimum value of routing hop-count $H_{mh}$, we need to consider following two cases. The first case is that the end-to-end routing hop-count of $P_s$ is $\mathbb{H}_h$ more than $P_{mh}$ in horizontal direction, and $\mathbb{H}_v$ less than $P_{mh}$ in vertical direction. If we represent the length of intra-orbit links as $L_v$ and the minimum length of inter-orbit links as $L_{h\text{min}}$, then the sufficient conditions for $H_s= H_{mh}$ in this case are
\begin{equation}
\mathbb{H}_hL_{h\text{min}} > \mathbb{H}_vL_v \label{C1}.
\end{equation}
\begin{equation}
\label{S1}
\begin{split}
s.t.\quad  \left\{\begin{array}{lc}
\mathbb{H}_h > \mathbb{H}_v, \\
\mathbb{H}_h \in \big[0, \lfloor \frac{N}{2} \rfloor \big], \vspace{1ex}\\
\mathbb{H}_v \in \big[0, \mathcal{Z}\big].\\
\end{array}\right.
\end{split}
\end{equation}
\begin{equation}
\mathcal{Z} = \begin{cases}
\text{mod}(F,M),&{\text{if $\text{mod}\big(F,M\big) \leq \big \lfloor \frac{M}{2} \big \rfloor$,}} \\
 M - \text{mod}(F,M),&{\text{if $\text{mod}\big(F,M\big) > \big \lfloor \frac{M}{2} \big \rfloor$.}}
\end{cases}
\end{equation}

Utilizing \eqref{S1}, we make a further derivation of \eqref{C1}, which can simplify \eqref{C1} and \eqref{S1} as follows.
\begin{equation}
\label{C11}
L_{h\text{min}} > \frac{\mathbb{H}_v^{\text{max}}}{\mathbb{H}_v^{\text{max}}+1} L_v.
\end{equation}
\begin{equation}
\label{S11}
\mathbb{H}_v^{\text{max}} = \text{min}\Big\{\mathcal{Z}, \big \lfloor \frac{N}{2} \big \rfloor - 1\Big\}.
\end{equation}

Moreover, the second case is that the end-to-end routing hop-count of $P_s$ is $\mathbb{H}_h$ less than $P_{mh}$
in horizontal direction, and $\mathbb{H}_v$ more than $P_{mh}$ in vertical direction. So in this case, the sufficient conditions
for $H_s= H_{mh}$ are
\begin{equation}
\mathbb{H}_vL_v >  \mathbb{H}_hL_{h\text{min}}. \label{C2}
\end{equation}
\begin{equation}
\label{S2}
\begin{split}
s.t.\quad  \left\{\begin{array}{lc}
\mathbb{H}_v > \mathbb{H}_h, \\
\mathbb{H}_h \in \big[0, \lfloor \frac{N}{2} \rfloor \big],  \vspace{1ex}\\
\mathbb{H}_v \in \big[0, \mathcal{Z}\big].\\
\end{array}\right.
\end{split}
\end{equation}

Similar to the first case, conditions \eqref{C2} and \eqref{S2} obtained in the second case can be further deduced as follows.
\begin{equation}
\label{C22}
L_v > \frac{\mathbb{H}_h^{\text{max}}}{\mathbb{H}_h^{\text{max}}+1}L_{h\text{min}}.
\end{equation}
\begin{equation}
\label{S22}
\mathbb{H}_h^{\text{max}} = \text{min}\Big\{\mathcal{Z}-1, \big \lfloor \frac{N}{2} \big \rfloor \Big\}.
\end{equation}

In conclusion, if the parameters of Walker Delta constellations can satisfy both \eqref{C11}, \eqref{S11} and \eqref{C22}, \eqref{S22}, the end-to-end
hop-count of the shortest distance path will be equal to the minimum value. Meanwhile, $L_{h\text{min}}$ and $L_v$ in above equations can be calculated as follows.
\begin{equation}
L_{h\text{min}} = 2\cdot( r_e+h)\cdot \text{sin}(\frac{\Gamma_\text{min}}{2}), \; L_v = 2\cdot\text{sin}(\frac{\pi}{M})\cdot(r_e+h).
\end{equation}
\begin{equation}
\begin{cases}
\Gamma_\text{min} = \text{arccos}\Big( \frac{1-\text{cos}\gamma}{2} - \frac{1+\text{cos}\gamma}{2}\text{cos}\big( 2\kappa - \Delta f\big)\Big),\\
\gamma = \text{arccos}\Big( 1- \frac{\big( \text{sin}\alpha \text{sin} \frac{2\pi}{N}\big)^2}{1+\text{cos}\frac{2\pi}{N}} \Big),\\
\kappa = \text{arcsin} \Big( \sqrt{\frac{\big(1+\text{cos} \frac{2\pi}{N}\big)^2}{2+2\text{cos}\frac{2\pi}{N} - \big( \text{sin}\alpha \text{sin} \frac{2\pi}{N}\big)^2}}\Big).\\
\end{cases}
\label{MinhopShort}
\end{equation}
\subsection{{Packet Format Design for Practical Implementation}}
\begin{figure}[htpb]
\centering
\includegraphics[width=0.45\textwidth]{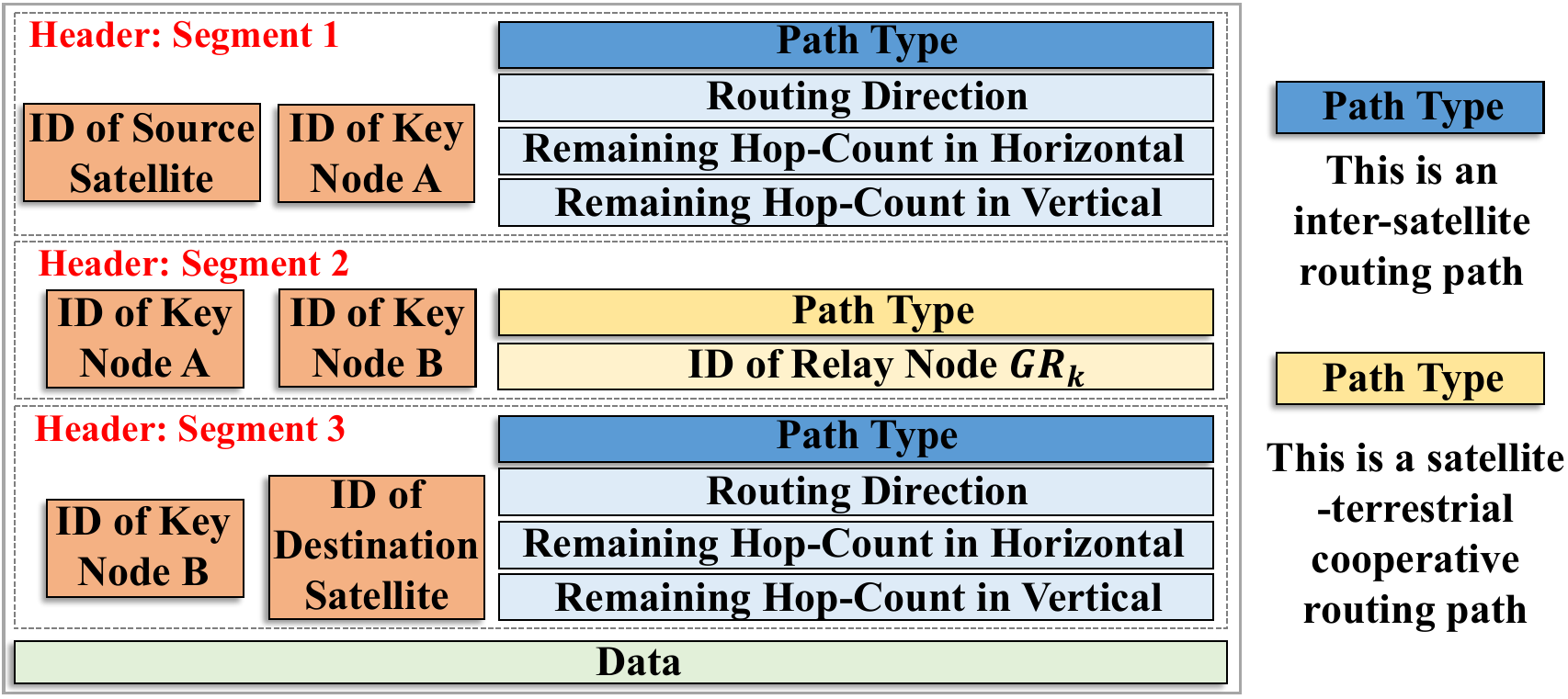}
\caption{The packet format in our proposal.}\label{figformat}
\end{figure}
To reduce routing overhead, we employ source routing paradigm, executing KNBG-MHCE only at source satellite and encapsulating its results into packets as shown in Fig.~\ref{figformat}. Subsequent nodes can make distributed routing decisions by extracting the encapsulated information, which includes both header and data. The header is structured into multiple segments, with each segment containing neighboring key nodes ID and the corresponding path information between them.
Taking Fig.~\ref{figformat} as an example, the path in segment1 is an inter-satellite routing path and this path is from source satellite to key node A. The path information of this path encompasses routing direction, remaining routing hop-count in both horizontal and vertical directions. When a node forwards this packet in a specific direction, the remaining routing hop-count in the corresponding direction within this packet should be subtracted by 1.
The path in segment2 represents a satellite-terrestrial cooperative routing path, and its path information includes the ID of utilized ground relay node $GR_k$. Furthermore, according to Fig.~\ref{figformat}, encapsulated segments within the packet header require updates in the following cases, and we use $\mathbb{A}$,$\mathbb{B}$ to denote key node A, B for brevity.
\begin{itemize}
\item \textbf{Case 1: }When $\mathbb{A}$ decides to forward this packet to ground relay $GR_k$, $\mathbb{A}$ removes segment1.
\item \textbf{Case 2: }The load status of downlink SGL between $\mathbb{A}$ and $GR_k$ exceeds the threshold $\Psi_{down}$ and $\mathbb{A}$ decides to forward this packet to another satellite gateway $SG_{k,j}$. At this point, $\mathbb{A}$ removes the original segment1 and executes Algorithm 1 using the coordinates of $\mathbb{A}$ and $SG_{k,j}$ as inputs. The results of Algorithm 1 are then encapsulated into this packet, formatted as new segment1. Meanwhile, the ID of key node A in segment2 is updated to the ID of $SG_{k,j}$.
\item \textbf{Case 3: }Due to satellite movement, $\mathbb{A}$ is no longer a key node. In response, $\mathbb{A}$ removes all segments within this packet, executes KNBG-MHCE with the IDs of $\mathbb{A}$ and destination satellite as inputs, and subsequently re-encapsulates the results of KNBG-MHCE into this packet.
\item \textbf{Case 4: }When this packet is forwarded to $\mathbb{B}$, $\mathbb{B}$ removes segment2.
\item \textbf{Case 5: }If the load status of uplink SGL between $GR_k$ and $\mathbb{B}$ exceeds the threshold $\Psi_{up}$, or if $\mathbb{B}$ is no longer a key node due to satellite movement, $GR_k$ will decide to forward this packet to another satellite gateway $SG_{k,q}$. At this point, $GR_k$ will first forward this packet to $SG_{k,q}$. $SG_{k,q}$ will then removes all segments within this packet, and executes KNBG-MHCE with the IDs of $SG_{k,q}$ and destination satellite as inputs, then re-encapsulates the results of KNBG-MHCE into this packet.
\end{itemize}

\section{Experiments and Performance Evaluation}\label{sec:sim}
\subsection{Simulation Setup}
In this section, extensive experiments are conducted under Starlink phase I constellation, and $N_0$ in Section \ref{sec:ours} is set to 4. Key node based graphs are regenerated every 600ms, and all simulations last 20.51s with the time interval $\tau$ of 1 ms\cite{TLR2014}. Moreover, we compare our proposal with other two types of strategies. Type I strategies only use ISLs for data forwarding, which consist of DSP, DIS\cite{DIS2022}, and LSP\cite{LSP2019}. Type II strategies include DSPCR, DISCR, LSPCR and our proposal, and the first three strategies in Type II are adaptations of Type I strategies for satellite-terrestrial cooperative routing scenario. Additionally, we deploy a total of 25 ground relays in two phases in the satellite-terrestrial cooperative routing scenario. 12 ground relays are deployed in the first phase, while other 13 ground relays are further added in the second phase, and the locations of all 25 ground relays are shown in Fig.~\ref{GRs}. In Fig.~\ref{GRs}, 12 ground relays in the first phase are represented by unfilled circles and 13 ground relays in the second phase are represented by color-filled circles. At the same time, the different colors indicate that the ground relays belong to different continents.
\begin{figure}[H]
\centering
\includegraphics[width = 0.8\linewidth]{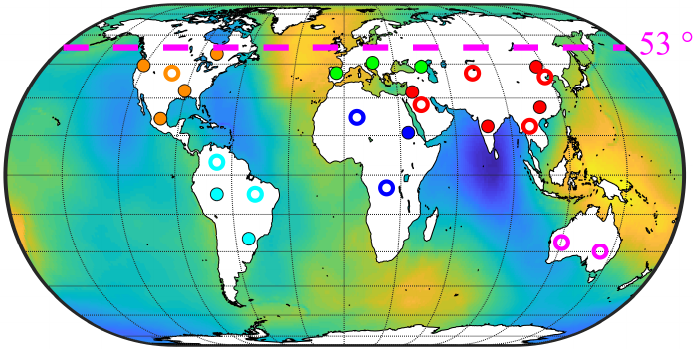}
\caption{The locations of all 25 ground relays.}
\label{GRs}
\end{figure}

Moreover, the overall simulation in this section consists of three parts. First, we evaluate the accuracy of Algorithm 1 and Algorithm 2, which are designed to estimate the minimum end-to-end hop-count for inter-satellite and satellite-terrestrial cooperative routing, respectively. Second, we discuss the impact of packet arrival rate and the size of satellite sending buffer queue on system throughput. Finally, we compare our proposal with other strategies in terms of routing hop-count, end-to-end delay, packet drop rate and system throughput. The key simulation parameters are shown in Table \ref{tabpara}.

\begin{table}[htpb]
\centering \caption{Simulation Parameters}
\label{tabpara}
\begin{tabular}{l l} \toprule 
\textbf{Parameter} & \textbf{Value} \\
\midrule 
\scriptsize  Configuration of constellations & \scriptsize  $1584/72/39/550/53^\circ$\\
\scriptsize  The size of sending buffer queue $B_s$ & \scriptsize  5Mbit, Ka / 10Mbit, laser\\
\scriptsize  The size of public waiting buffer queue $B_w$ & \scriptsize  40Mbit, Ka / 1Gbit, laser\\
\scriptsize  The rate of ISL $R_{ISL}$ & \scriptsize  25Mbps, Ka / 2.5Gbps, laser\\
\scriptsize The rate of SGL $R_{SGL}$ in Ka band & \scriptsize 1.5Gbps downlink / 2Gbps uplink\\
\scriptsize The rate of SGL $R_{SGL}$ in laser band & \scriptsize 5Gbps downlink / 5Gbps uplink\\
\scriptsize The packet arrival rate $R_{pac}$ under Ka ISLs &  \scriptsize 1-9Mbps\\
\scriptsize The packet arrival rate $R_{pac}$ under laser ISLs &  \scriptsize 80-350Mbps\\
\scriptsize Smoothing factors $\eta_1 - \eta_5$&  \scriptsize (0.01, Ka / 0.5, laser), 2, 1, 2, 1\\
\bottomrule 
\end{tabular}
\end{table}
\subsection{The Accuracy Assessment of Designed Minimum End-to-End Hop-Count Estimation Methods}
\begin{figure}[htpb]
\centering
\includegraphics[width=0.34\textwidth]{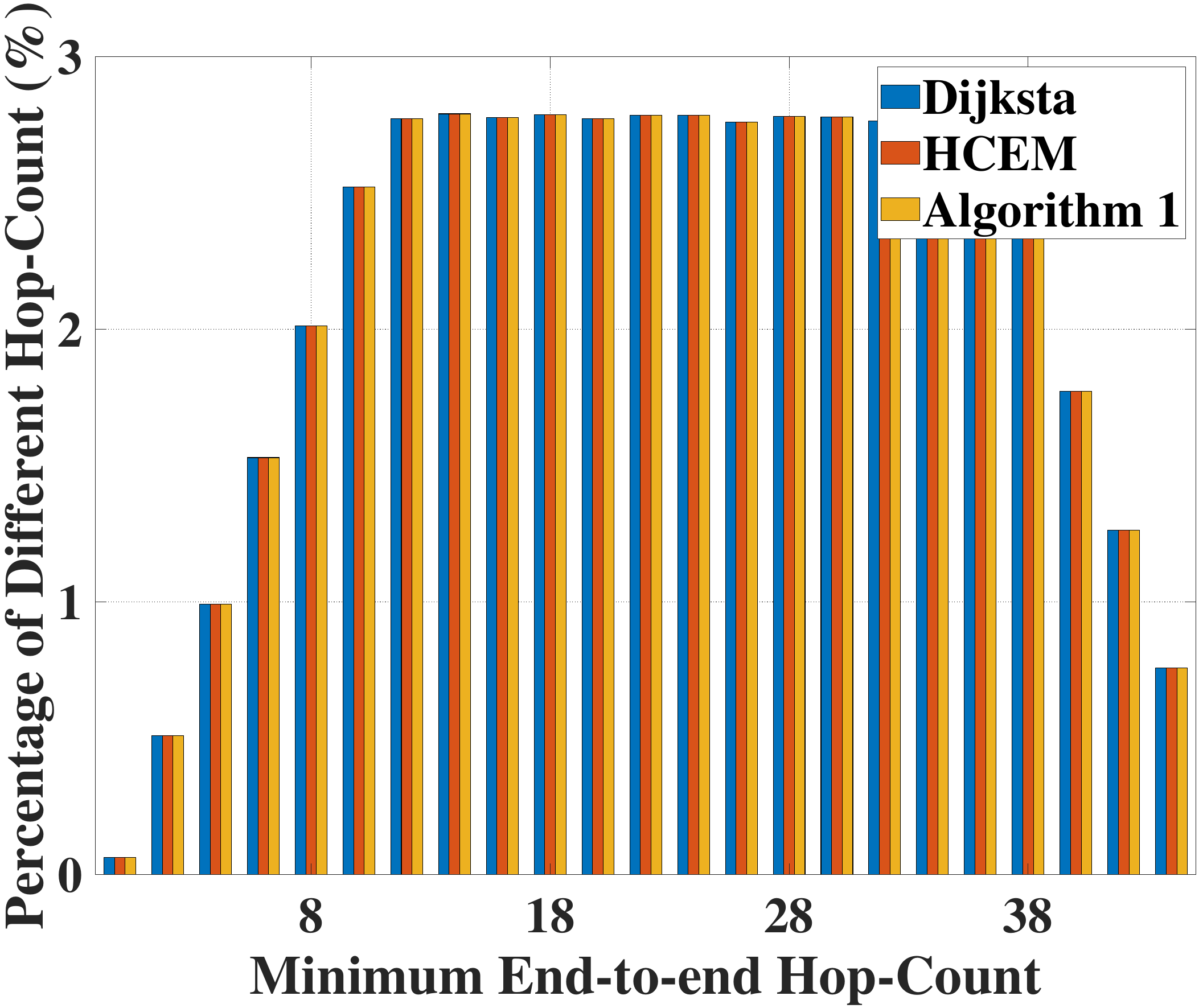}
\caption{The accuracy assessment of Algorithm 1.}\label{fig16}
\end{figure}

\begin{figure}[htpb]
\centering
\includegraphics[width=0.36\textwidth]{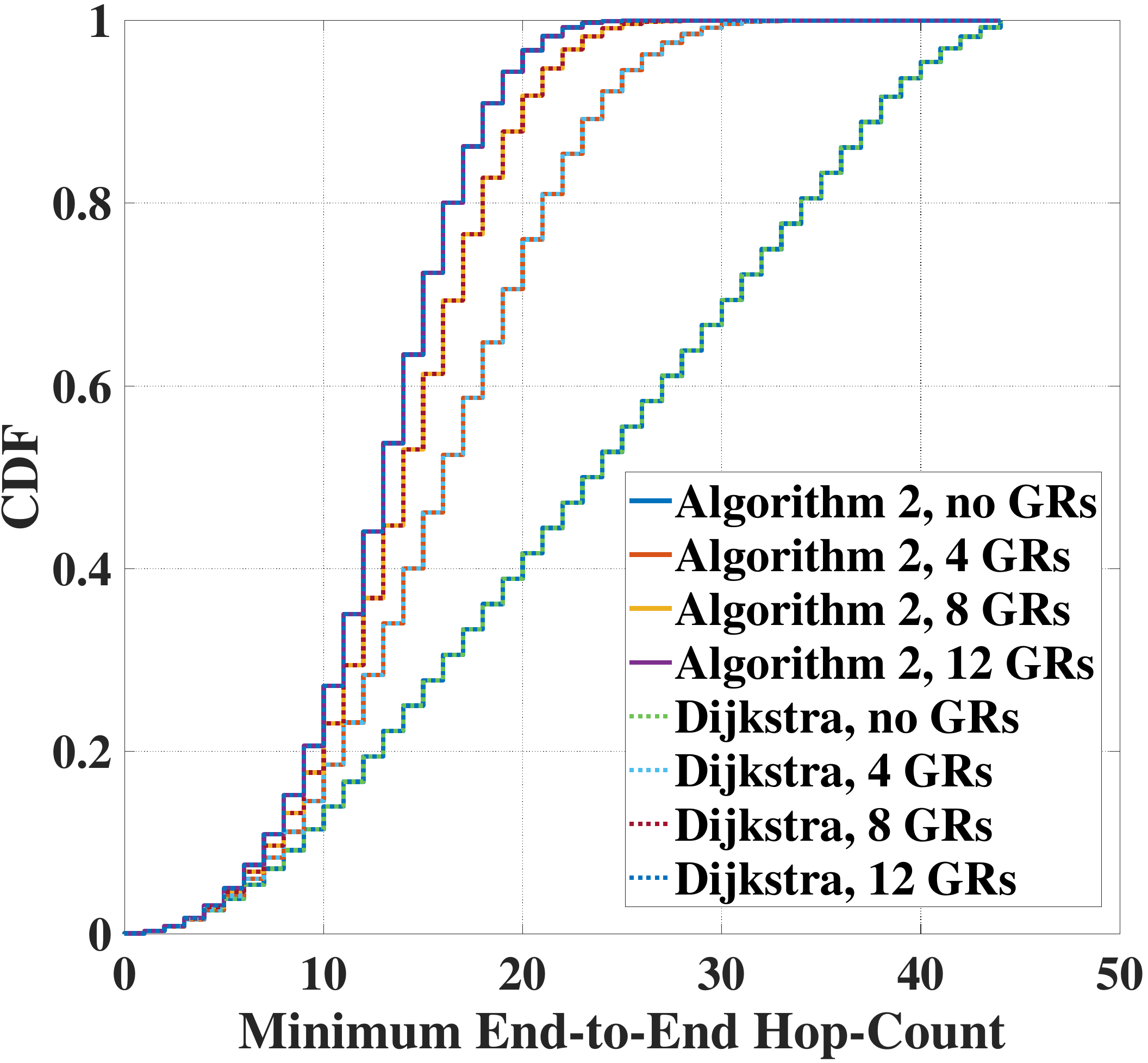}
\caption{The accuracy assessment of Algorithm 2.}\label{fig17}
\end{figure}

Considering there are a total of 1584 satellites in Starlink phase I constellation, we number them from 1 to 1584 and generate 3 million random numbers, where the odd numbers are used as source satellite nodes and the even numbers are used as destination satellite nodes. As a result, 1.5 million source-destination pairs are set to estimate the minimum end-to-end hop-count with Algorithm 1, Dijkstra and Hop-Count Estimation Method (HCEM)\cite{MinHop2021}, respectively. Fig.~\ref{fig16} illustrates that the percentage of different minimum hop-count obtained by Algorithm 1 is consistent with the other two methods, indicating that the designed Algorithm 1 can accurately estimate the minimum end-to-end hop-count for inter-satellite forwarding mode.

Furthermore, we verify the accuracy of KNBG-MHCE (Algorithm 2) under different numbers of ground relays (GRs). As shown in Fig.~\ref{fig17}, the CDF curves of minimum end-to-end hop-count obtained by KNBG-MHCE overlap with those obtained by Dijkstra, reviewing that the designed KNBG-MHCE can accurately estimate the minimum end-to-end hop-count for satellite-terrestrial cooperative routing scenarios.
\subsection{The Evaluation of Average System Throughput Under Different $R_{pac}$ and $B_s$}\label{TwoPar}
To fully discuss the impact of packet arrival rate $R_{pac}$ and the size of sending buffer queue $B_s$ on average system throughput, we conduct simulations in system with Ka band ISLs and laser band ISLs, respectively. In both systems, we set 3000 non-persistent on-off flows similar to\cite{ELB2008}, and employ Ka band SGLs along with our proposal as uniform routing strategy. Notably, average throughput here refers to the average size of packets reaching destination nodes per unit of time, and it can be further defined as follows \cite{help3}
\begin{equation}
\centering
\small \text{Throughput} = \mathop{\text{lim}}\limits_{T\rightarrow \infty} \frac{\sum_{t=0}^T\text{Packets reaching destinations}}{T}.
\end{equation}
Fig.~\ref{fig18} and Fig.~\ref{fig19} represent the variation of average throughput in system with Ka and laser band ISLs, respectively. Under same parameters, we can observe that the system with laser band ISLs always exhibits a higher throughput compared to the system with Ka band ISLs. This is because that the transmission rate will rise as $R_{ISL}$ increases, which results in the enhancement in system throughput.
\begin{figure}[htpb]
\centering
\includegraphics[width=0.4\textwidth]{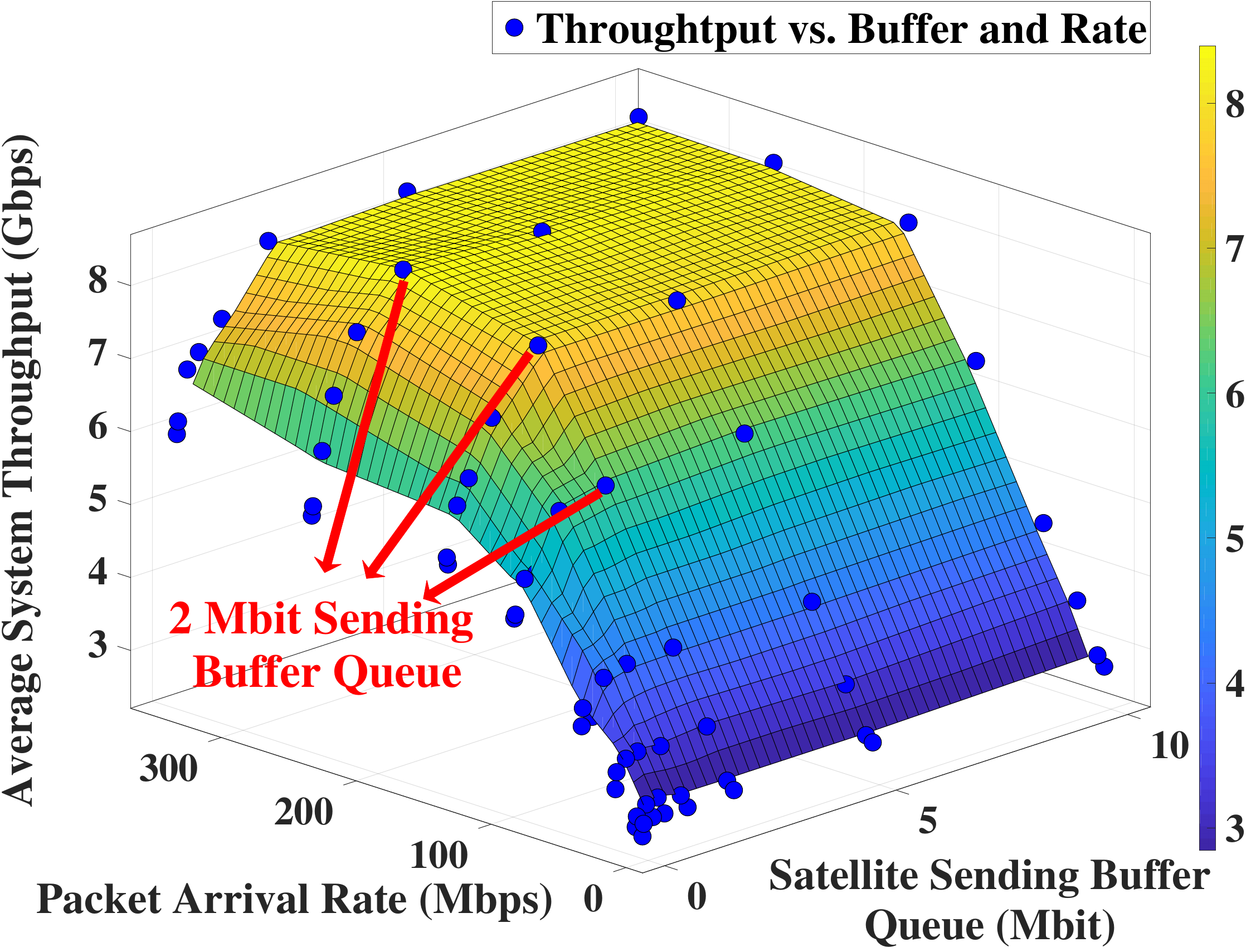}
\caption{Average system throughput with Ka band ISLs.\label{fig18}}
\end{figure}

\begin{figure}[htpb]
\centering
\includegraphics[width=0.4\textwidth]{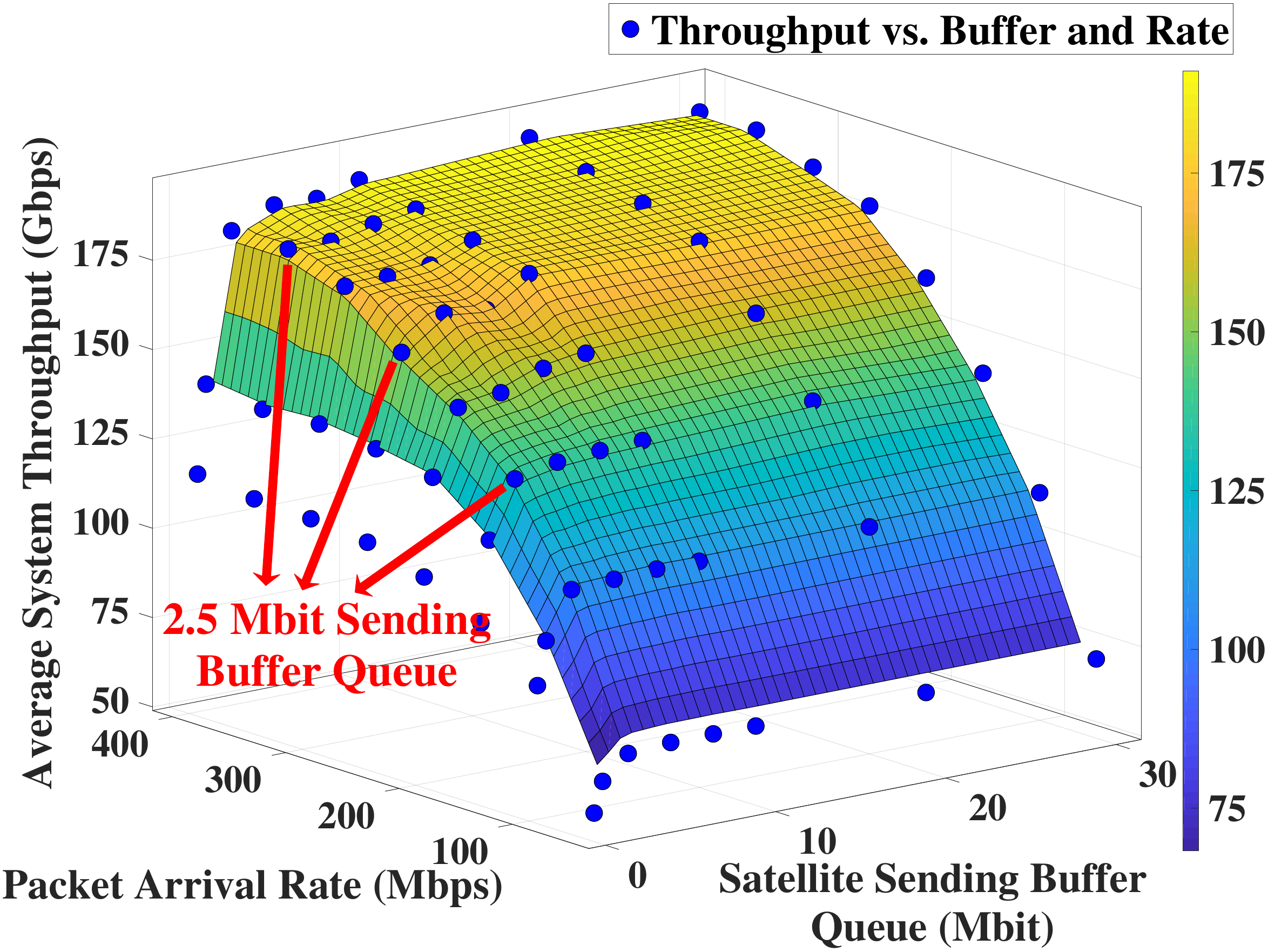}
\caption{Average system throughput with laser band ISLs.\label{fig19}}
\end{figure}
{Moreover, when $R_{pac}$ is fixed, the average system throughput will increase till $B_s$ reaches a threshold $B_{s}^{th}$ in both figures above. $B_s^{th}$ is determined by the maximum transmission rate per unit time of all links.} In Fig.~\ref{fig18}, the maximum transmission rate in Ka band system is $2\text{Gbps}\times 1\text{ms} = 2\text{Mbit}$. Therefore, when $B_s \geq 2\text{Mbit}$, the size of packets that each link can store reaches the upper limit of its transmission capacity. As a result, the throughput remains almost unchanged regardless of further variations in $B_s$. Similarly in Fig.~\ref{fig19}, since the maximum data transmission rate is $2.5\text{Gbps}\times 1\text{ms} = 2.5\text{Mbit}$, $B_s^{th}$ becomes 2.5 Mbit in laser system.
On the other hand, when $B_s$ is fixed, in both Fig.~\ref{fig18} and Fig.~\ref{fig19}, the average system throughput improves with the increase of $R_{pac}$. However, when $R_{pac}$ reaches a threshold $R_{pac}^{th}$, system will enter saturated state and the throughput remains relatively stable thereafter. From Fig.~\ref{fig18} we can see that $R_{pac}^{th}$ is about 150Mbps in the system with Ka band ISLs, and with the increase of link rate, $R_{pac}^{th}$ grows to 350Mbps in system with laser band ISLs, which is shown in Fig.~\ref{fig19}.

 It's worth noting that system should reach saturation state when packet arrival rate exceeds the lowest transmission rate of all links. However, as shown in Fig.~\ref{fig18}, $R_{pac}^{th}$ in system with Ka band ISLs is 150Mbps, which is much larger than the lowest transmission rate $R_{ISL} = 25$Mbps. This happens because {some source nodes can reach their destinations just through SGLs, and no longer need to go through the low-rate ISLs.} Therefore, system will enter saturation state only when the transmission capacity of SGL is fully utilized, which leads to the increase in $R^{th}_{pac}$ to 150Mbps.
\subsection{Performance Comparison among Routing Strategies}
{We conduct simulations of different routing strategies in two systems as before, and the related results are shown in Fig.~\ref{Kasim} and Fig.~\ref{lasersim}, respectively. In the following figures, end-to-end delay refers to the sum of propagation delay and queuing delay. Packet forwarding rate refers to the number of packets forwarded by a satellite per unit of time, which is determined by the satellite processing rate and link transmission rate. Additionally, for more precise results, we delete the data flow which can be transmitted from source to destination just by SGLs as mentioned in Section \ref{TwoPar}.}
 \begin{figure*}[htpb]
\centering
\subfigure[]{
\begin{minipage}[t]{0.25\linewidth} 
\centering
\includegraphics[width = 0.93\linewidth]{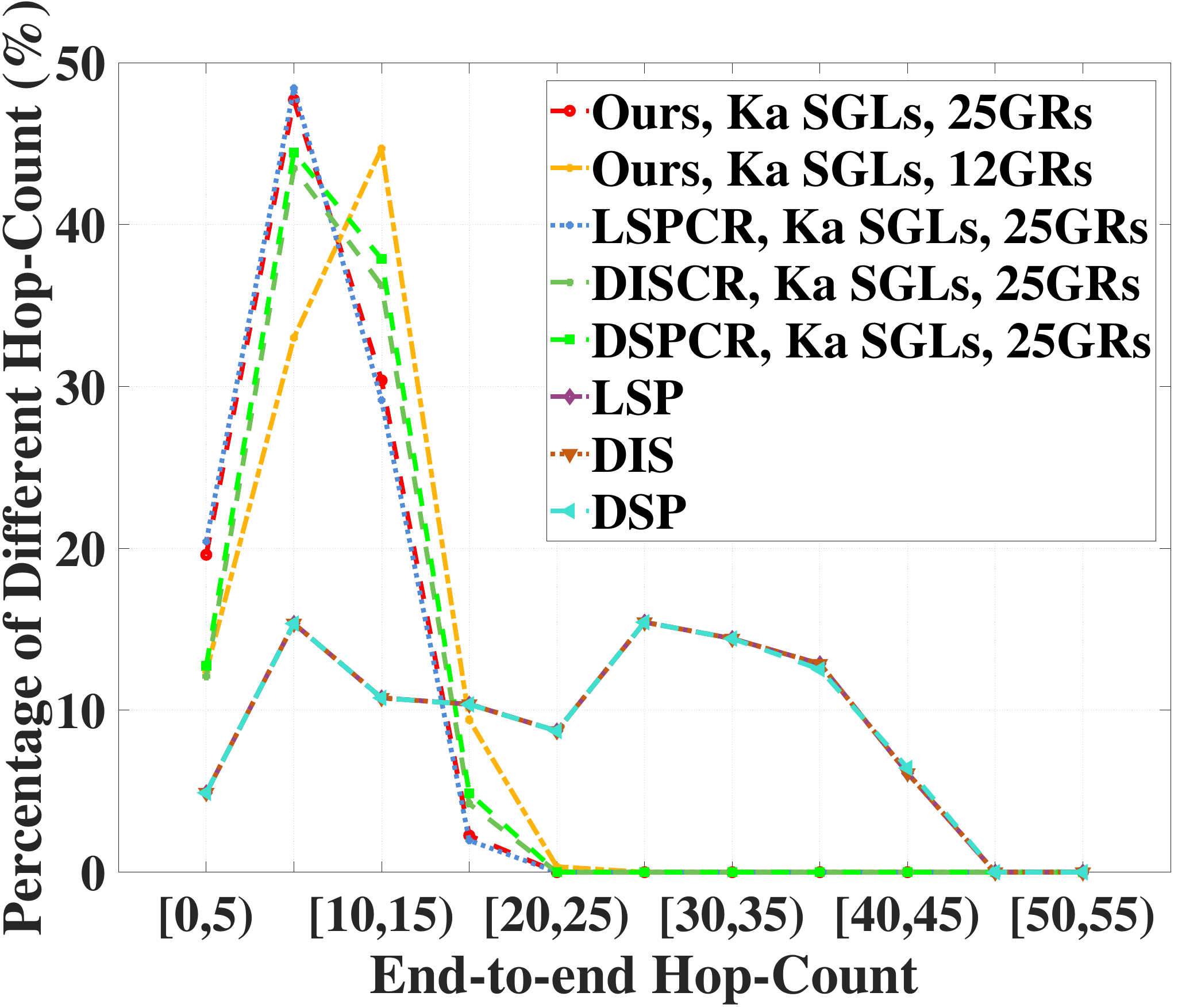}
\label{fig32}
\end{minipage}%
}%
\subfigure[]{
\begin{minipage}[t]{0.25\linewidth} 
\centering
\includegraphics[width = 0.99\linewidth]{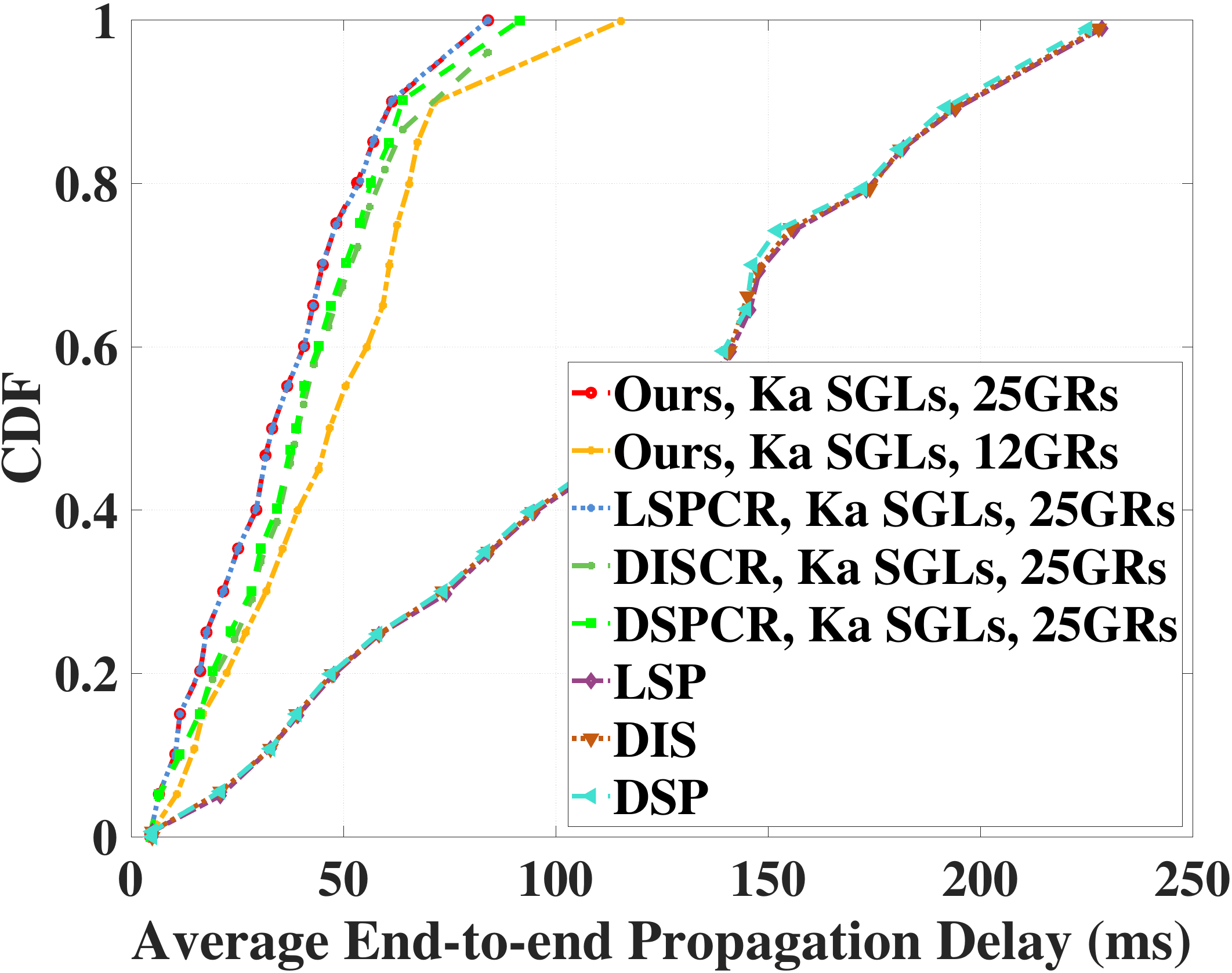}
\label{fig33}
\end{minipage}%
}%
\subfigure[]{
\begin{minipage}[t]{0.25\linewidth} 
\centering
\includegraphics[width = 0.97\linewidth]{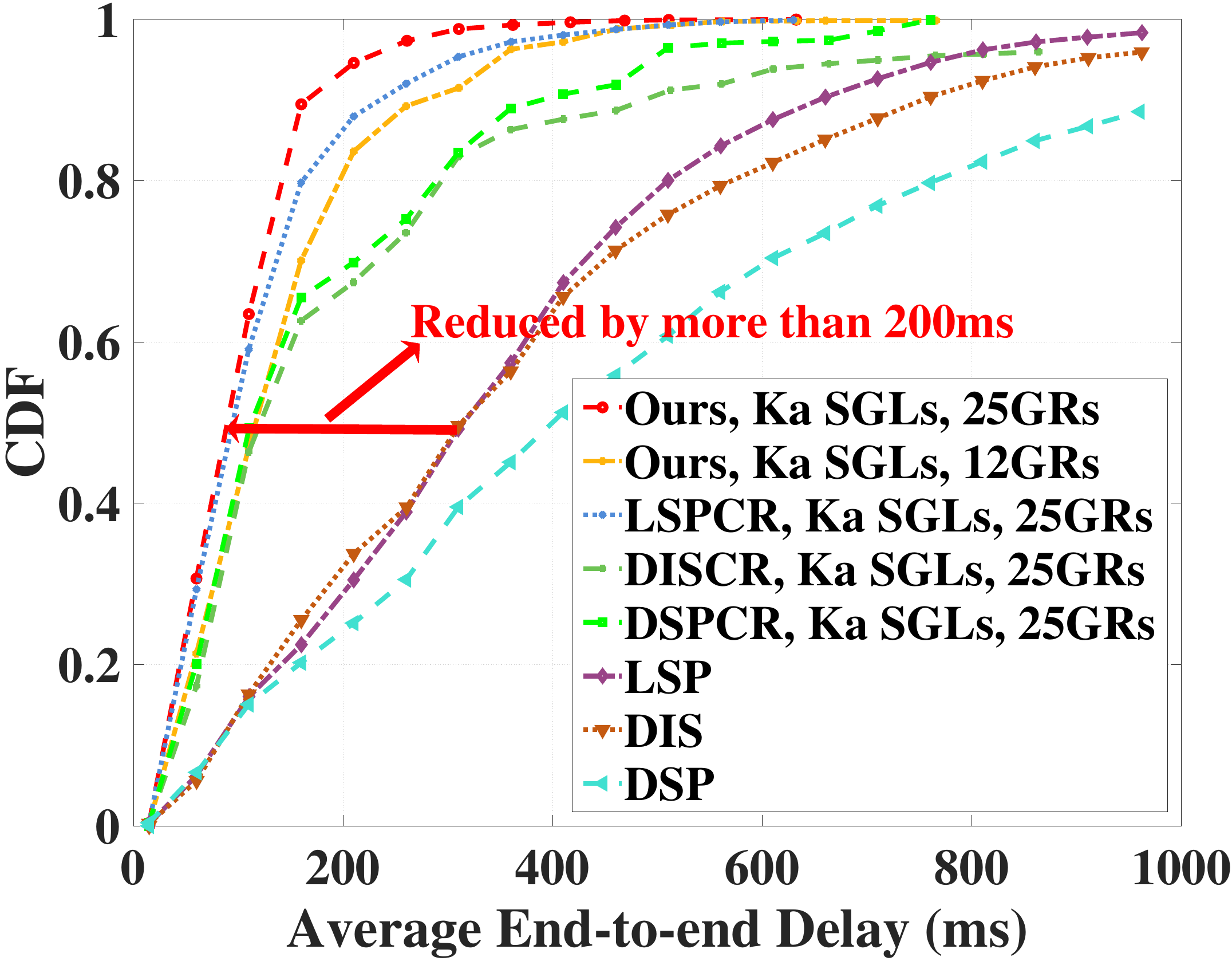}
\label{fig34}
\end{minipage}%
}%
\subfigure[]{
\begin{minipage}[t]{0.25\linewidth} 
\centering
\includegraphics[width = 0.97\linewidth]{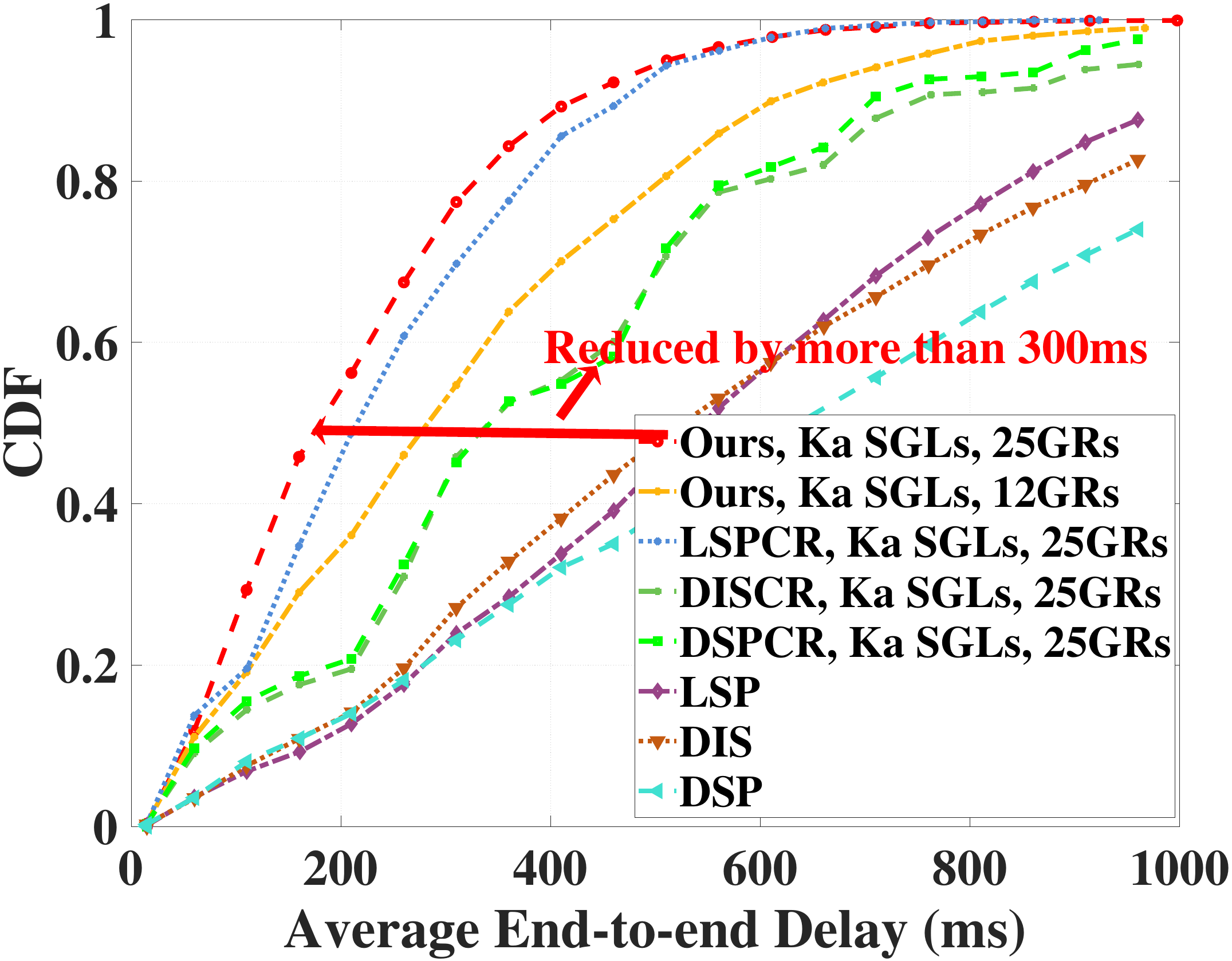}
\label{fig35}
\end{minipage}%
}%
\qquad
\subfigure[]{
\begin{minipage}[t]{0.25\linewidth} 
\centering
\includegraphics[width = 0.93\linewidth]{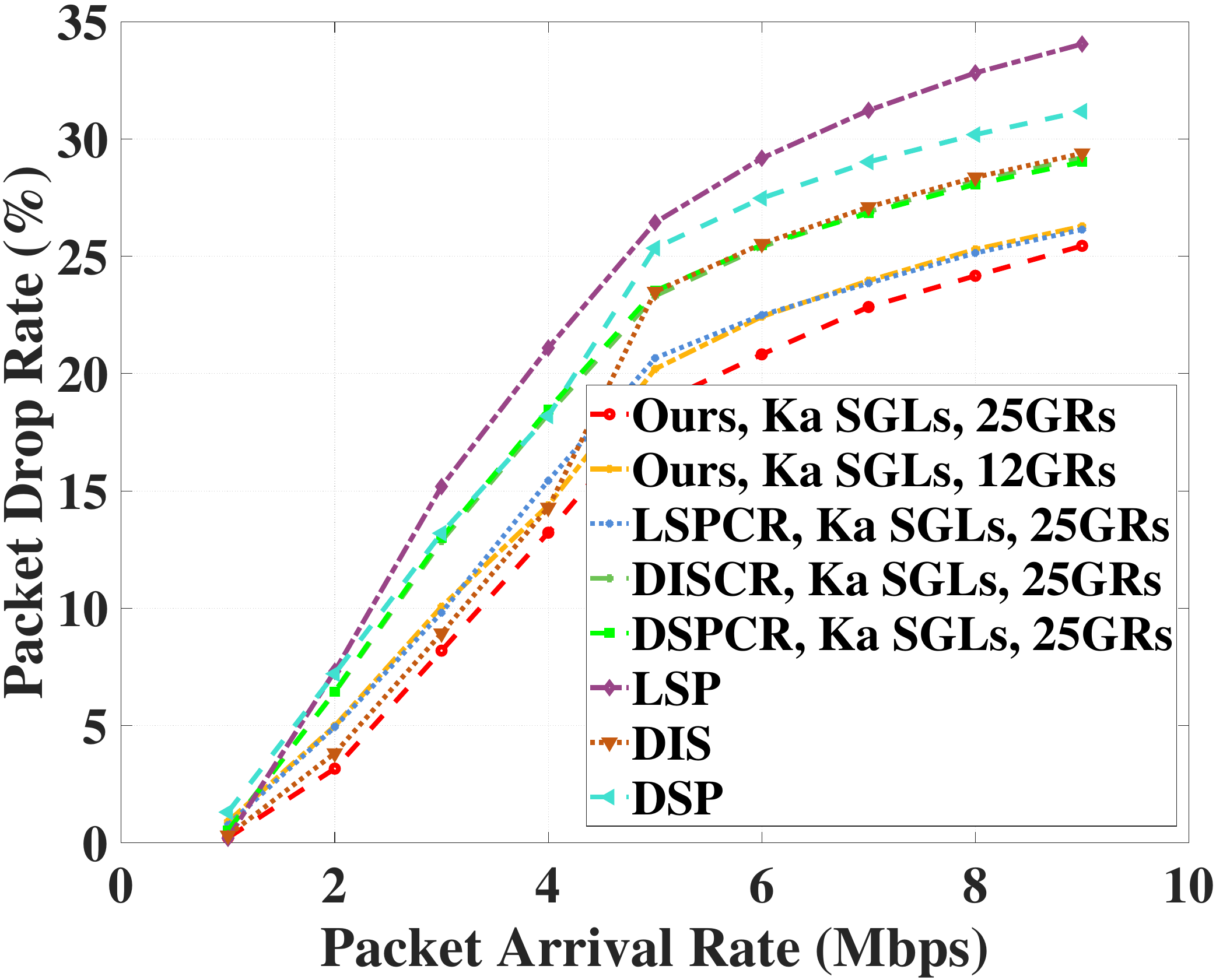}
\label{fig29}
\end{minipage}%
}%
\subfigure[]{
\begin{minipage}[t]{0.25\linewidth} 
\centering
\includegraphics[width = 0.9\linewidth]{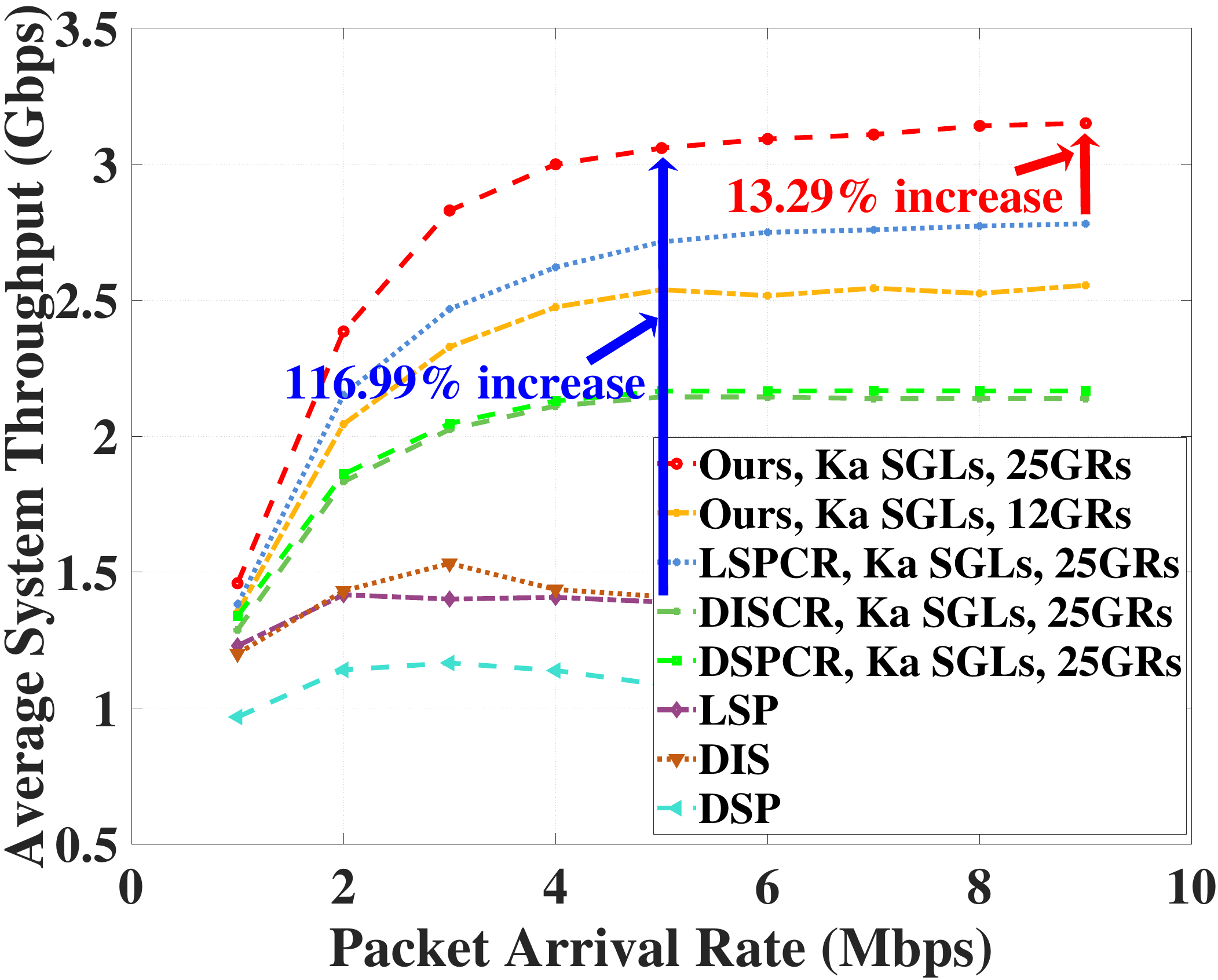}
\label{fig28}
\end{minipage}%
}%
\subfigure[]{
\begin{minipage}[t]{0.25\linewidth} 
\centering
\includegraphics[width = 1\linewidth]{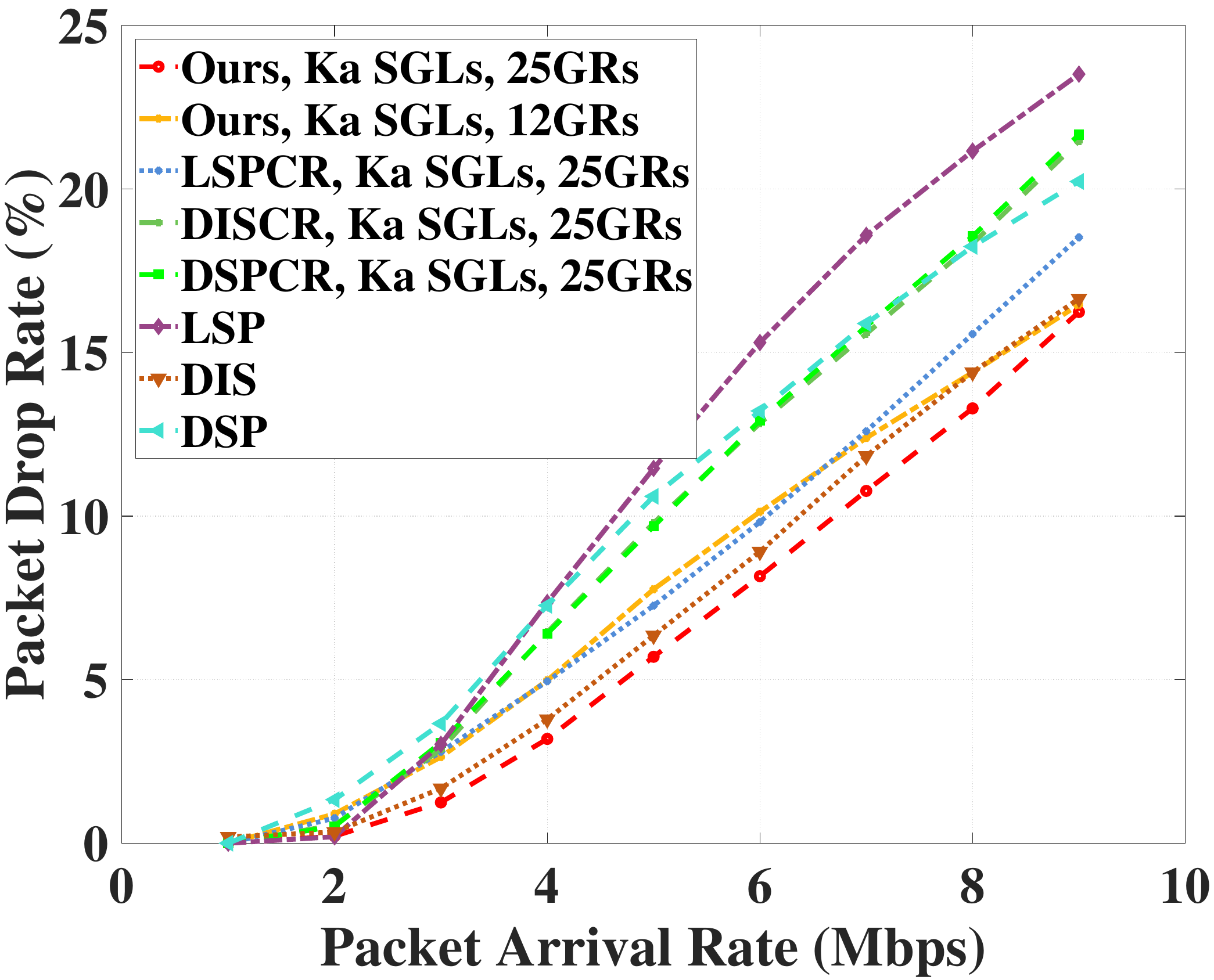}
\label{fig31}
\end{minipage}%
}%
\subfigure[]{
\begin{minipage}[t]{0.25\linewidth} 
\centering
\includegraphics[width = 1\linewidth]{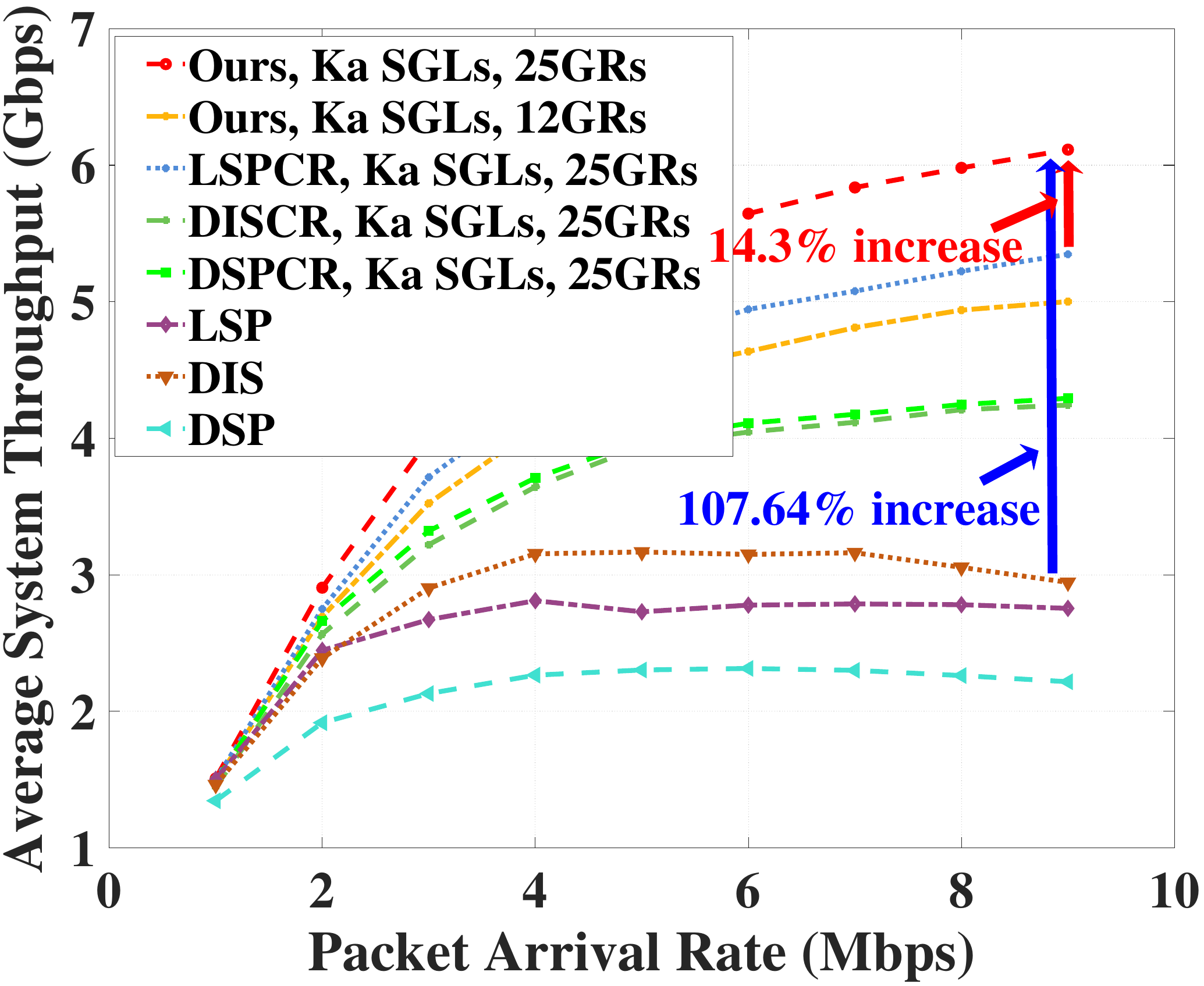}
\label{fig30}
\end{minipage}%
}
\centering
\caption{The simulation results with ISLs in Ka band, where (a) is the distribution of end-to-end routing hop-count, (b) illustrates the CDF of end-to-end propagation delay, (c) presents the CDF of average end-to-end delay with $R_{pac} = 1\text{Mbps}$, (d) presents the CDF of average end-to-end delay with $R_{pac} = 4\text{Mbps}$, (e) shows the packet drop rate with different $R_{pac}$, (f) shows the average system throughput with different $R_{pac}$, (g) depicts the packet drop rate at double packet forwarding rate with different $R_{pac}$, (h) depicts the average system throughput at double packet forwarding rate with different $R_{pac}$.}
 \label{Kasim}
\end{figure*}
\begin{figure*}[htpb]
\centering
\subfigure[]{
\begin{minipage}[t]{0.25\linewidth} 
\centering
\includegraphics[width = 0.93\linewidth]{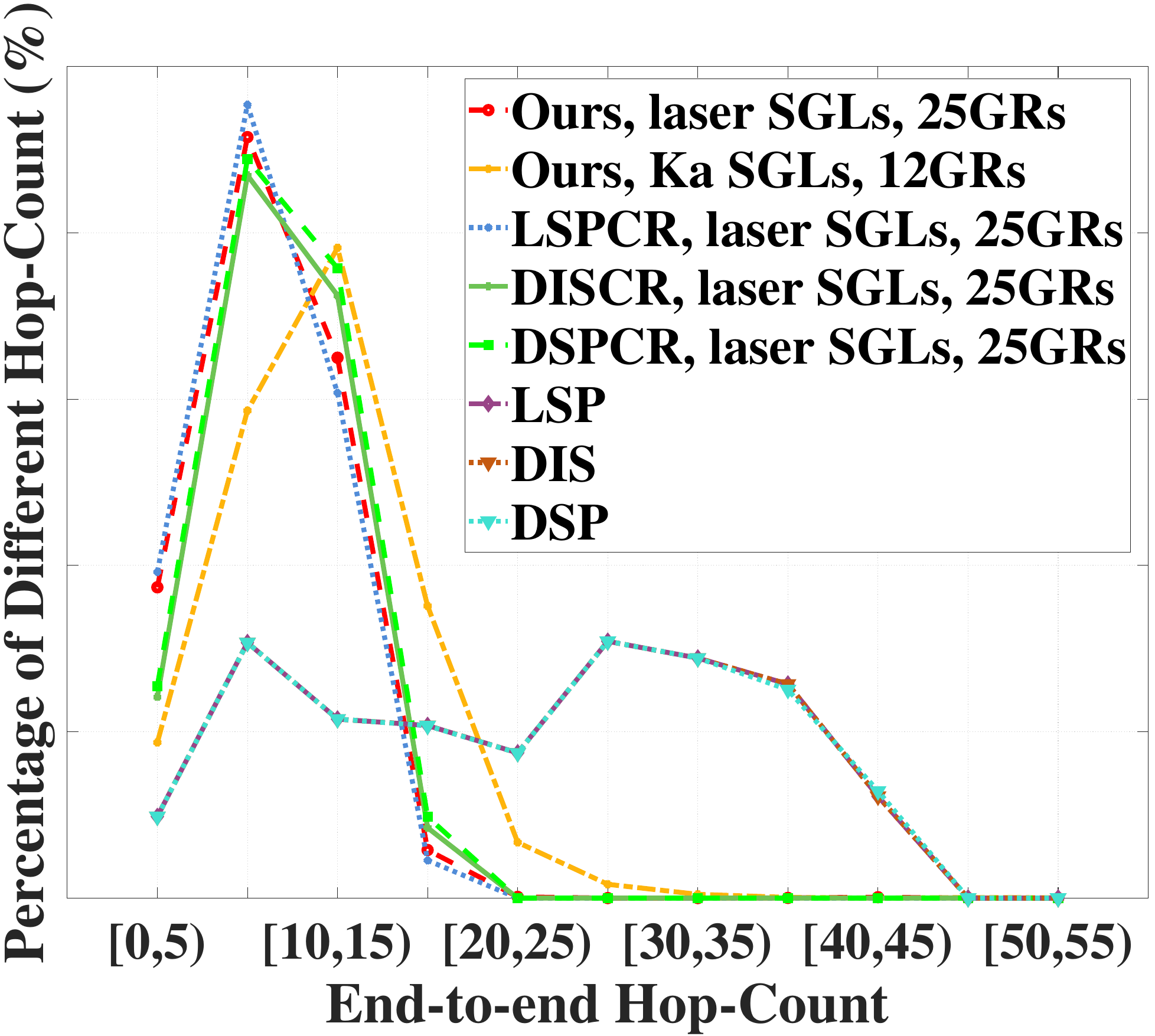}
\label{fig24}
\end{minipage}%
}%
\subfigure[]{
\begin{minipage}[t]{0.25\linewidth} 
\centering
\includegraphics[width = 1\linewidth]{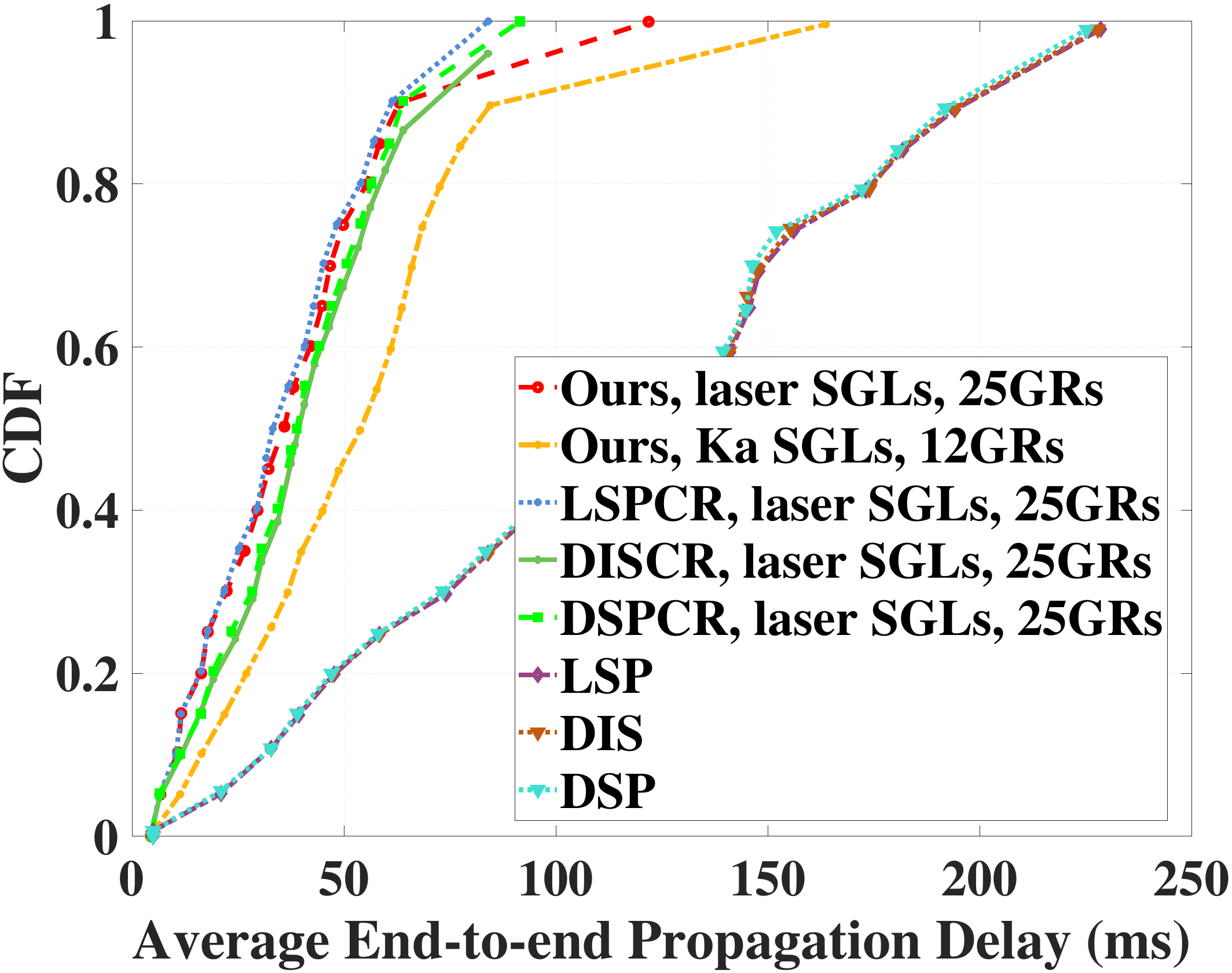}
\label{fig25}
\end{minipage}%
}%
\subfigure[]{
\begin{minipage}[t]{0.25\linewidth} 
\centering
\includegraphics[width = 0.97\linewidth]{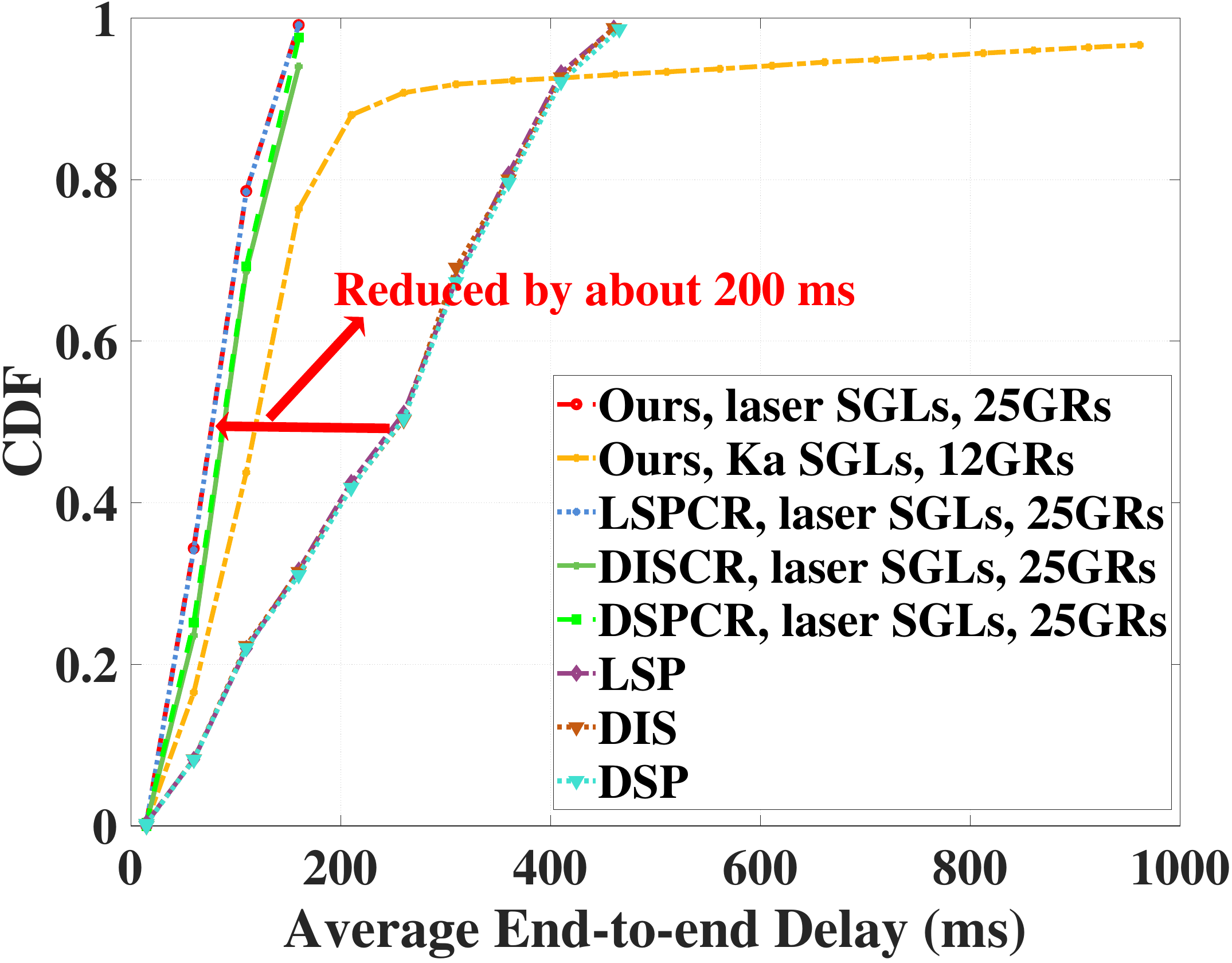}
\label{fig26}
\end{minipage}%
}%
\subfigure[]{
\begin{minipage}[t]{0.25\linewidth} 
\centering
\includegraphics[width = 0.97\linewidth]{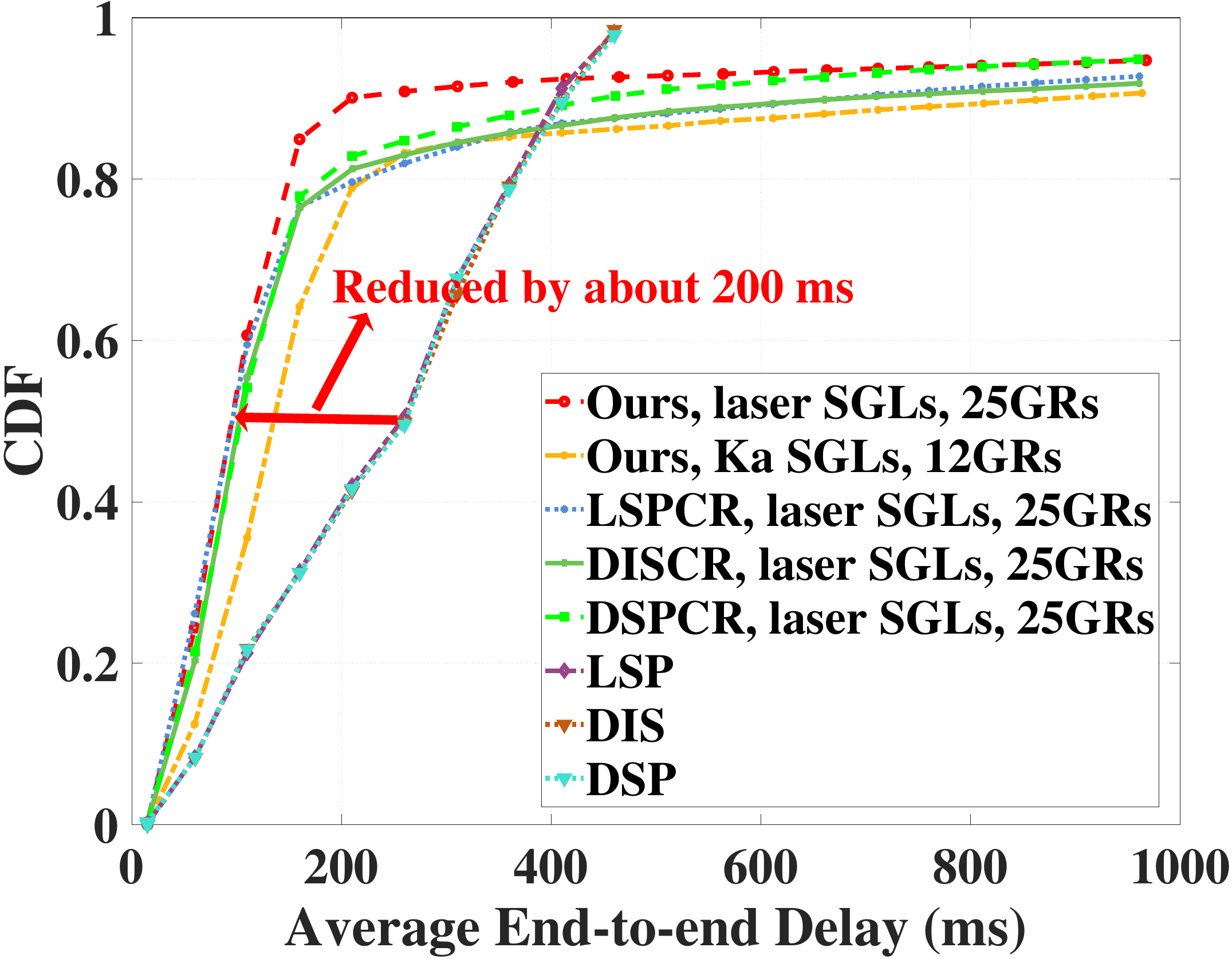}
\label{fig27}
\end{minipage}%
}%
\quad
\subfigure[]{
\begin{minipage}[t]{0.25\linewidth} 
\centering
\includegraphics[width = 0.93\linewidth]{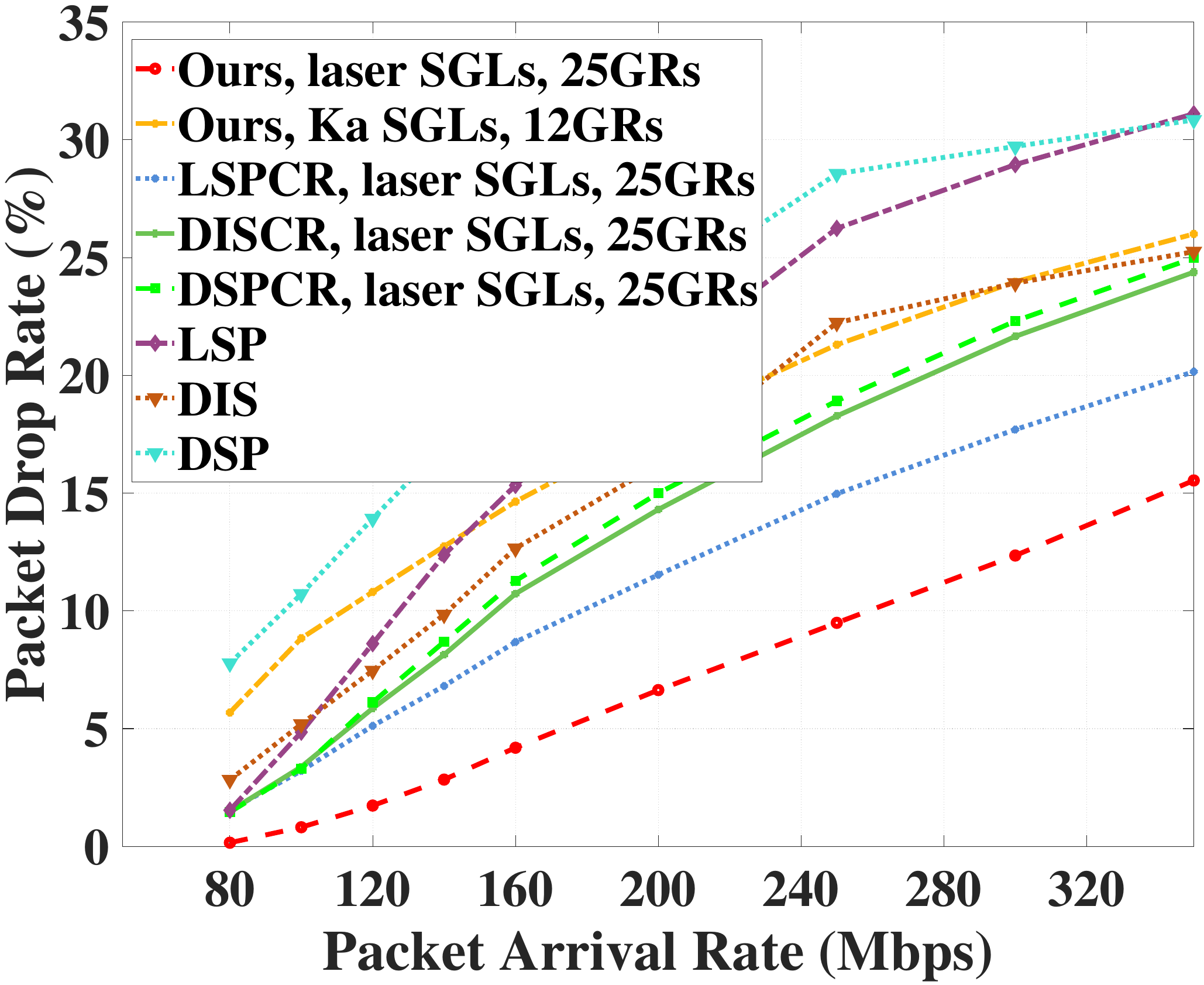}
\label{fig21}
\end{minipage}%
}%
\subfigure[]{
\begin{minipage}[t]{0.25\linewidth} 
\centering
\includegraphics[width = 0.93\linewidth]{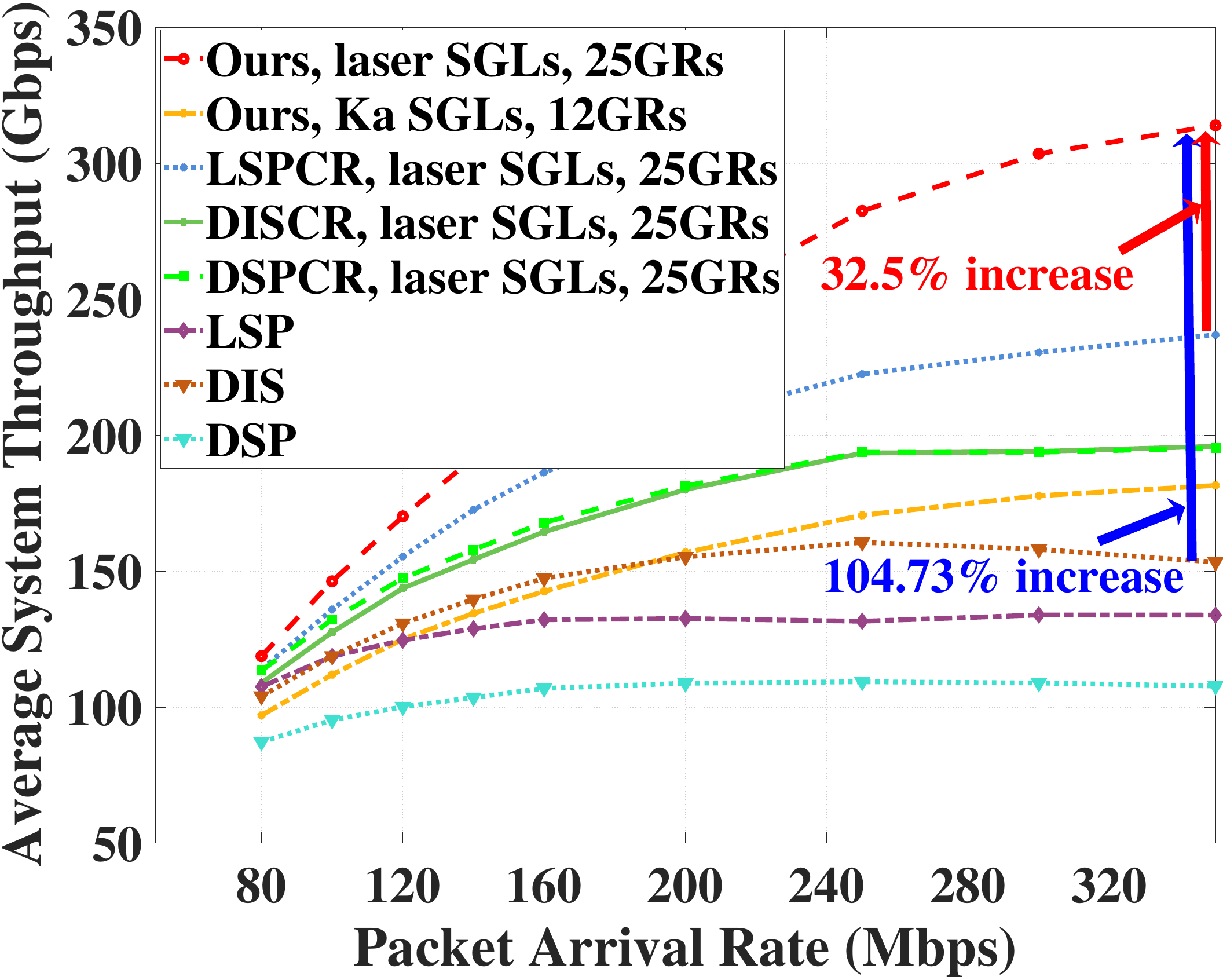}
\label{fig20}
\end{minipage}%
}%
\subfigure[]{
\begin{minipage}[t]{0.25\linewidth} 
\centering
\includegraphics[width = 1\linewidth]{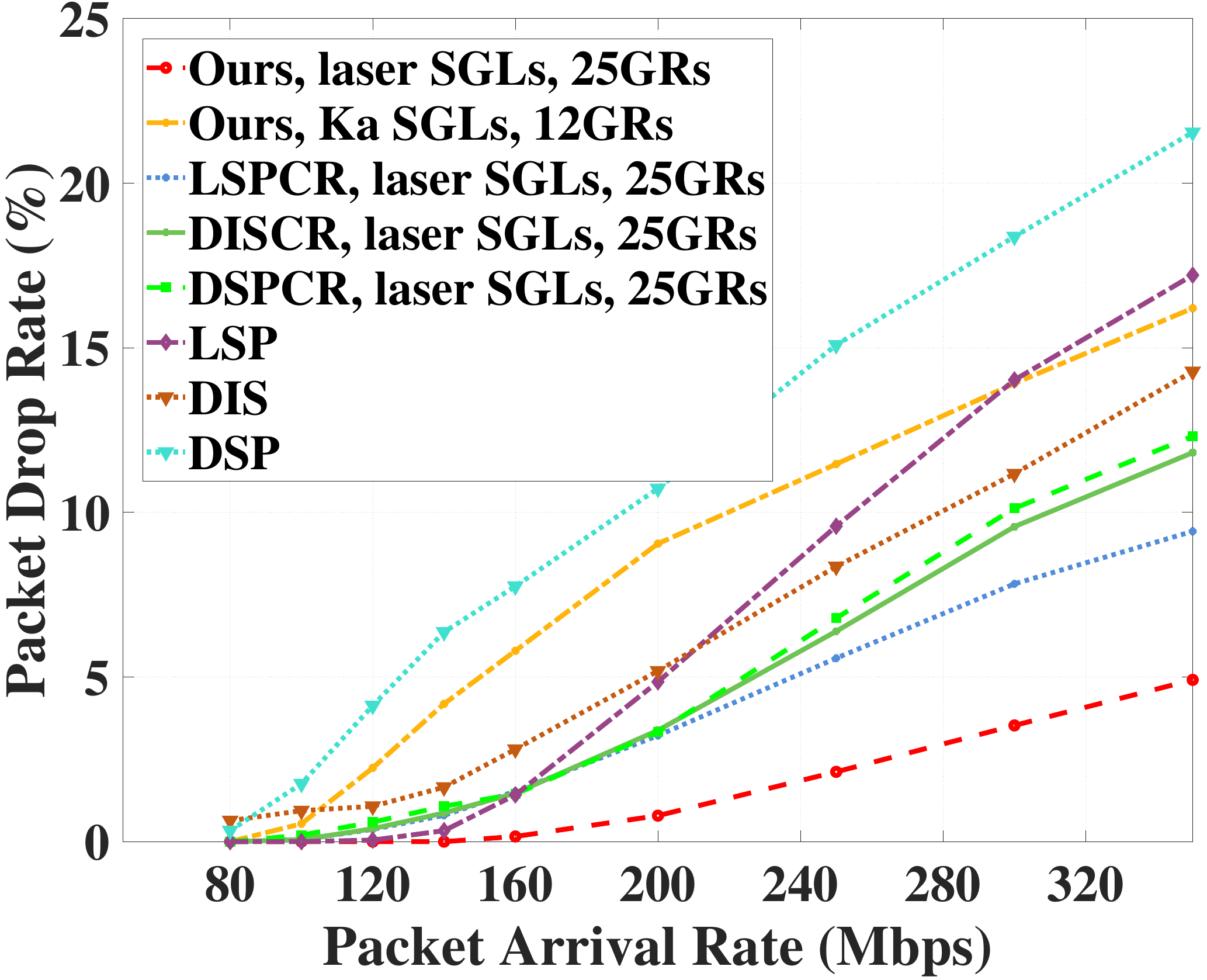}
\label{fig23}
\end{minipage}%
}%
\subfigure[]{
\begin{minipage}[t]{0.25\linewidth} 
\centering
\includegraphics[width = 1\linewidth]{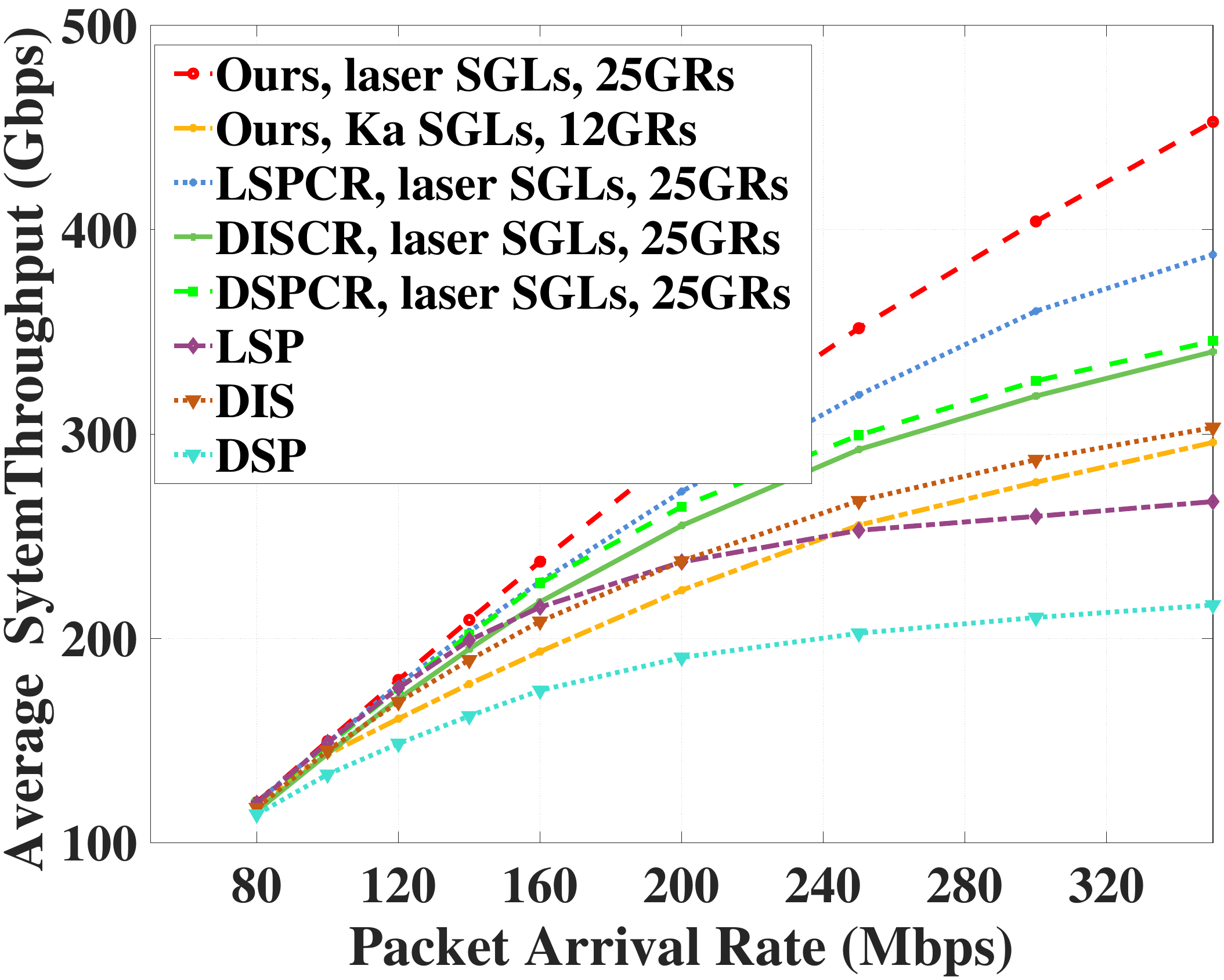}
\label{fig22}
\end{minipage}%
}%
\centering
\caption{The simulation results with ISLs in laser band, where (a) is the distribution of end-to-end routing hop-count, (b) illustrates the CDF of end-to-end propagation delay, (c) presents the CDF of average end-to-end delay with $R_{pac} = 80\text{Mbps}$, (d) presents the CDF of average end-to-end delay with $R_{pac} = 350\text{Mbps}$, (e) shows the packet drop rate with different $R_{pac}$, (f) shows the average system throughput with different $R_{pac}$, (g) depicts the packet drop rate at double packet forwarding rate with different $R_{pac}$, (h) depicts the average system throughput at double packet forwarding rate with different $R_{pac}$.}
\label{lasersim}
\end{figure*}
\subsubsection{End-to-end Hop-Count}
Fig.~\ref{fig32} and Fig.\ref{fig24} represent the end-to-end hop-count distribution under different strategies in two systems. Since the end-to-end routing hop-count is not related to the frequency of ISLs, the trends of curves in these two figures are essentially the same. Taking Fig.~\ref{fig32} as an example, under an equivalent number of ground relays, Dijkstra-based routing strategies exhibit hop-count distributions that are nearly identical to strategies based on minimum end-to-end hop-count. This is because that  the end to-end routing hops of the shortest distance path tends to be the minimum value in mega constellations, which is fully discussed in Section \ref{hopcount}. Moreover, the average hop-count of Type II strategies is notably lower than that of Type I strategies, and the more ground relays there are, the smaller the average end-to-end hop-count of Type II strategies is, which further proves that satellite-terrestrial cooperative routing can significantly reduce the forwarding hops. {Meanwhile, the cumulative distribution of end-to-end propagation delay under different strategies in the two systems are shown as Fig.~\ref{fig33} and \ref{fig25}. Since the end-to-end propagation delay under mega constellations is closely related to the number of routing hops, the average propagation delay of Type II strategies is much smaller than that of Type I strategies in both figures as well.}
\subsubsection{End-to-end Delay}
We analyze the end-to-end delay with packet arrival rate of initial rate $R_{pac}^{0}$ and the threshold $R_{pac}^{th}$.
Fig.~\ref{fig34} and \ref{fig35} depict the cumulative distributions of end-to-end delay under two different rates in system with Ka band ISLs, while Fig.~\ref{fig26} and \ref{fig27} show that in system with laser band ISLs. It can be observed that the average end-to-end delay increases with $R_{pac}$ in both systems, and comparing with Type I strategies, our proposal can reduce the average end-to-end delay by more than 200ms in all cases.
\subsubsection{Packet Drop Rate and Average System Throughput}
Packet drop rate is the ratio of the total number of packets dropped by system to the total number of packets generated in system during the simulation period, and it can be further defined as \cite{help3}
\begin{equation}
\centering
\text{Packet Drop Rate} = \frac{\text{Number of dropped packets}}{\text{Number of generated packets}}.
\end{equation}
Fig.~\ref{fig29} and Fig.~\ref{fig21} show the variation of packet drop rate in the two systems, and it can be observed that our proposal always has the smallest packet drop rate.  Consequently, on account of the the lowest end-to-end delay and the smallest packet drop rate, our proposal markedly enhances system throughput as shown in Fig.~\ref{fig28} and Fig.~\ref{fig20}. Comparing with Type I strategies, our proposal can improve system throughput by more than 116.99\% in system with Ka band ISLs and over 104.73\% in system with laser band ISLs. Furthermore, in contrast to Type II strategies, our proposal can improve system throughput with the increase of  13.29\% and 32.5\%, respectively.
\subsubsection{Packet forwarding rate}
{Fig.~\ref{fig31} and Fig.~\ref{fig30} represent the packet drop rate and average throughput of system with Ka band ISLs at  double packet forwarding rate, while Fig.~\ref{fig23} and Fig.~\ref{fig22} illustrate the relevant performances of the system with laser band ISLs. Comparing Fig.~\ref{fig31} with Fig.~\ref{fig29} (or Fig.~\ref{fig23} with Fig.~\ref{fig21}), it can be found that packet drop rate will significantly decrease at the same $R_{pac}$. This is because that as packet forwarding rate increases, packets can be forwarded more efficiently, and the packets stored in each node become less, which leads to the reduction in packet drop rate. Moreover, as the packet drop rate reduces, average throughput in both systems will achieve a significant improvement. Comparing Fig.~\ref{fig30} and Fig.~\ref{fig28}, it can be reviewed that when the packet forwarding rate is increased by a factor of two, the upper bound of system throughput will be boosted by nearly 200\% as well.}
\section{Conclusions}\label{sec:con}
In this paper, to address the challenge of large end-to-end forwarding hops in mega constellations, we have proposed a distributed satellite-terrestrial cooperative routing approach by jointly considering the minimum end-to-end hop-count and queuing delay constraints. Particularly, to achieve an accurate and low-complexity minimum end-to-end hop-count estimation, we have first introduced a geographic region division method of the earth's surface and realized the transformation of 3D constellation into a 2D graph RTPG by calculating in which region each satellite is located. Next, by extracting key nodes in RTPG, we create the key node based graph KNBG. Utilizing KNBG as input, a low complexity minimum end-to-end hop-count estimation method KNBG-MHCE has been designed for satellite-terrestrial cooperative routing. Meanwhile, we have analyzed some properties of our proposal and designed the packet format for practical implementation. Finally, based on simulation results, we have discussed the relationship between system throughput and parameters such as sending buffer queue and packet arrival rate. It has also been verified that our proposal can significantly reduce the end-to-end delay and improve system throughput compared to the other two types of routing strategies.

Owing to the superior performance of our proposal, the proposed routing strategy can be used to guide packet forwarding in future mega LEO satellite constellations such as Starlink. Under these mega constellations, our proposal can not only significantly reduce the end-to-end routing hop-count, lower end-to-end delay, and improve system throughput, but also ensure low complexity and overhead in routing table construction. Based on the source routing paradigm and the designed packet format, our proposal can be naturally converged with IP protocol by extending the header of IP packet, which brings convenience for practical implementation. In addition, using medium earth orbit (MEO) satellites for cooperative routing to reduce the end-to-end routing hop-count is also a promising area, and this will be conducted in our future work.
\bibliographystyle{IEEEtran}

\begin{thebibliography}{99}
\bibitem{sun2022}
Y.~Sun, M.~Peng, S.~Zhang \emph{et~al.}, ``Integrated satellite-terrestrial
  networks: architectures, key techniques, and experimental progress,''
  \emph{IEEE Netw.}, vol.~36, no.~6, pp. 191--198, Dec. 2022.

\bibitem{tao}
H.~Xu \emph{et~al.}, ``Joint beam scheduling and beamforming design for
  cooperative positioning in multi-beam leo satellite networks,'' \emph{IEEE
  Trans. Veh. Technol.}, 2023, doi: 10.1109/TVT.2023.3332142.

\bibitem{TMC1}
M.~Hu, M.~Xiao, Y.~Hu \emph{et~al.}, ``Software defined multicast using segment
  routing in leo satellite networks,'' \emph{IEEE Trans Mob Comput}, vol.~23,
  no.~1, pp. 835--849, Oct. 2022.

\bibitem{help1}
H.~Zhang, C.~Jiang, J.~Wang \emph{et~al.}, ``Multicast beamforming optimization
  in cloud-based heterogeneous terrestrial and satellite networks,'' \emph{IEEE
  Trans. Veh. Technol.}, vol.~69, no.~2, pp. 1766--1776, Dec. 2019.

\bibitem{LEOnetwork2021}
I.~Leyva-Mayorga, B.~Soret, and P.~Popovski, ``Inter-plane inter-satellite
  connectivity in dense leo constellations,'' \emph{IEEE Trans. Wirel.
  Commun.}, vol.~20, no.~6, pp. 3430--3443, Jun. 2021.

\bibitem{help5}
Y.~Feng \emph{et~al.}, ``Performance analysis in satellite communication with
  beam hopping using discrete-time queueing theory,'' \emph{IEEE Internet
  Things J.}, Nov. 2023.

\bibitem{help4}
R.~Wang, M.~A. Kishk, and M.-S. Alouini, ``Stochastic geometry-based low
  latency routing in massive leo satellite networks,'' \emph{IEEE Trans Aerosp
  Electron Syst}, vol.~58, no.~5, pp. 3881--3894, Aug. 2022.

\bibitem{Timegraph2023}
N.~Zhang, Z.~Na, J.~Tao \emph{et~al.}, ``Time-varying graph and binary tree
  search based routing algorithm for leo satellite networks,'' \emph{IEEE
  Trans. Veh. Technol.}, pp. 1--6, 2023, doi:10.1109/TVT.2023.3274134.

\bibitem{multipath}
F.~Tang, H.~Zhang, and L.~T. Yang, ``Multipath cooperative routing with
  efficient acknowledgement for leo satellite networks,'' \emph{IEEE Trans Mob
  Comput}, vol.~18, no.~1, pp. 179--192, Jan. 2019.

\bibitem{yuan2023}
S.~Yuan \emph{et~al.}, ``Joint network function placement and routing
  optimization in dynamic software-defined satellite-terrestrial integrated
  networks,'' \emph{IEEE Trans. Wirel. Commun.}, pp. 1--15, Oct. 2023,
  doi:10.1109/TWC.2023.3324729.

\bibitem{chen2022delay}
L.~Chen \emph{et~al.}, ``Delay-optimal cooperation transmission in remote
  sensing satellite networks,'' \emph{IEEE Trans Mob Comput}, Sept. 2023.

\bibitem{DVTR1997}
M.~Werner, ``A dynamic routing concept for atm-based satellite personal
  communication networks,'' \emph{IEEE J. Sel. Areas Commun.}, vol.~15, no.~8,
  pp. 1636--1648, Oct. 1997.

\bibitem{VN1997}
R.~Mauger and C.~Rosenberg, ``Qos guarantees for multimedia services on a
  tdma-based satellite network,'' \emph{IEEE Commun Mag.}, vol.~35, no.~7, pp.
  56--65, Jul. 1997.

\bibitem{DISLOAD}
X.~Deng, L.~Chang, S.~Zeng \emph{et~al.}, ``Distance-based back-pressure
  routing for load-balancing leo satellite networks,'' \emph{IEEE Trans. Veh.
  Technol.}, vol.~72, no.~1, pp. 1240--1253, Sep. 2022.

\bibitem{time2022}
Z.~Han, C.~Xu, G.~Zhao \emph{et~al.}, ``Time-varying topology model for dynamic
  routing in leo satellite constellation networks,'' \emph{IEEE Trans. Veh.
  Technol.}, vol.~72, no.~3, pp. 3440--3454, Mar. 2023.

\bibitem{graph1}
F.~Li, S.~Chen, M.~Huang \emph{et~al.}, ``Reliable topology design in
  time-evolving delay-tolerant networks with unreliable links,'' \emph{IEEE
  Trans Mob Comput}, vol.~14, no.~6, pp. 1301--1314, Jun. 2015.

\bibitem{graph2}
M.~Huang, S.~Chen, Y.~Zhu \emph{et~al.}, ``Topology control for time-evolving
  and predictable delay-tolerant networks,'' \emph{IEEE Trans Comput}, vol.~62,
  no.~11, pp. 2308--2321, Nov. 2013.

\bibitem{graph3}
A.~Casteigts, P.~Flocchini, W.~Quattrociocchi \emph{et~al.}, ``Time-varying
  graphs and dynamic networks,'' in \emph{Ad-hoc, Mobile, and Wireless
  Networks: 10th International Conference, ADHOC-NOW 2011}, Paderborn, Germany,
  Jul. 2011, pp. 346--359.

\bibitem{Netgrid}
J.~Li, H.~Lu, K.~Xue \emph{et~al.}, ``Temporal netgrid model-based dynamic
  routing in large-scale small satellite networks,'' \emph{IEEE Trans. Veh.
  Technol.}, vol.~68, no.~6, pp. 6009--6021, Apr. 2019.

\bibitem{database}
J.~A. Fraire and E.~L. Gasparini, ``Centralized and decentralized routing
  solutions for present and future space information networks,'' \emph{IEEE
  Netw.}, vol.~35, no.~4, pp. 110--117, Aug. 2021.

\bibitem{flood1}
J.~Wang, F.~Xu, and F.~Sun, ``Benchmarkinng of routing protocols for layered
  satellite networks,'' in \emph{The Proceedings of the Multiconference on"
  Computational Engineering in Systems Applications"}, Beijing, China, Oct.
  2006, pp. 1087--1094.

\bibitem{GlobeDIS}
H.~Yang, B.~Guo, X.~Xue \emph{et~al.}, ``Interruption tolerance strategy for
  leo constellation with optical inter-satellite link,'' \emph{IEEE Trans.
  Netw. Service Manag}, 2023, doi:10.1109/TNSM.2023.3274638.

\bibitem{ELB2008}
T.~Taleb \emph{et~al.}, ``Explicit load balancing technique for ngeo satellite
  ip networks with on-board processing capabilities,'' \emph{IEEE ACM Trans
  Netw}, vol.~17, no.~1, pp. 281--293, Feb. 2009.

\bibitem{LSP2019}
Q.~Chen, X.~Chen, L.~Yang \emph{et~al.}, ``A distributed congestion avoidance
  routing algorithm in mega-constellation network with multi-gateway,''
  \emph{Acta Astronaut.}, vol. 162, pp. 376--387, May. 2019.

\bibitem{TLR2014}
G.~Song, M.~Chao, B.~Yang \emph{et~al.}, ``Tlr: A traffic-light-based
  intelligent routing strategy for ngeo satellite ip networks,'' \emph{IEEE
  Trans. Wirel. Commun.}, vol.~13, no.~6, pp. 3380--3393, Jun. 2014.

\bibitem{res1}
Y.~Lu, Y.~Zhao, F.~Sun \emph{et~al.}, ``Enhancing transmission efficiency of
  mega-constellation leo satellite networks,'' \emph{IEEE Trans. Veh.
  Technol.}, vol.~71, no.~12, pp. 13\,210--13\,225, Dec. 2022.

\bibitem{res2}
Z.~Lin, H.~Li, J.~Liu \emph{et~al.}, ``Inter-networking and function
  optimization for mega-constellations,'' in \emph{2022 IFIP Networking
  Conference (IFIP Networking)}, Catania, Italy, Jun. 2022, pp. 1--9.

\bibitem{res3}
G.~Fan, H.~Li, J.~Liu \emph{et~al.}, ``User-driven flexible and effective link
  connection design for mega-constellation satellite networks,'' in \emph{2023
  International Wireless Communications and Mobile Computing (IWCMC)},
  Marrakesh, Morocco, Jun. 2023, pp. 793--799.

\bibitem{res4}
H.~Yan, Q.~Zhang, and Y.~Sun, ``A novel routing scheme for leo satellite
  networks based on link state routing,'' in \emph{2014 IEEE 17th International
  Conference on Computational Science and Engineering}, Chengdu, China, Dec.
  2014, pp. 876--880.

\bibitem{res5}
H.~Li, H.~Zhang, L.~Qiao \emph{et~al.}, ``Queue state based dynamical routing
  for non-geostationary satellite networks,'' in \emph{2018 IEEE 32nd
  International Conference on Advanced Information Networking and Applications
  (AINA)}, Krakow, Poland, May. 2018, pp. 1--8.

\bibitem{res6}
H.~Nishiyama, Y.~Tada, N.~Kato \emph{et~al.}, ``Toward optimized traffic
  distribution for efficient network capacity utilization in two-layered
  satellite networks,'' \emph{IEEE Trans. Veh. Technol.}, vol.~62, no.~3, pp.
  1303--1313, Mar. 2012.

\bibitem{res7}
Q.~Chen, L.~Yang, X.~Liu \emph{et~al.}, ``Multiple gateway placement in
  large-scale constellation networks with inter-satellite links,'' \emph{INT J
  SATELL COMM N}, vol.~39, no.~1, pp. 47--64, Jan. 2021.

\bibitem{MinHop2021}
Q.~Chen, G.~Giambene, L.~Yang \emph{et~al.}, ``Analysis of inter-satellite link
  paths for leo mega-constellation networks,'' \emph{IEEE Trans. Veh.
  Technol.}, vol.~70, no.~3, pp. 2743--2755, Mar. 2021.

\bibitem{res9}
A.~U. Chaudhry and H.~Yanikomeroglu, ``Laser intersatellite links in a starlink
  constellation: A classification and analysis,'' \emph{IEEE Veh. Technol.
  Mag.}, vol.~16, no.~2, pp. 48--56, Jun. 2021.

\bibitem{res10}
M.~Handley, ``Delay is not an option: Low latency routing in space,'' in
  \emph{Proceedings of the 17th ACM Workshop on Hot Topics in Networks}, New
  York, USA, Nov. 2018, pp. 85--91.

\bibitem{res11}
J.~Iridium, ``Perfectly in sync, while traveling more than 30,000 kilometers
  per hour,'' 2022.

\bibitem{res12}
M.~Albulet, ``Spacex non-geostationary satellite system: Attachment a technical
  information to supplement schedules,'' \emph{US Fed. Commun. Comm.}, 2016.

\bibitem{res13}
Q.~Zhu, H.~Tao, Y.~Cao \emph{et~al.}, ``Laser inter-satellite link visibility
  and topology optimization for mega constellation,'' \emph{Electronics},
  vol.~11, no.~14, p. 2232, Jul. 2022.

\bibitem{res14}
Y.~Lee and J.~P. Choi, ``Connectivity analysis of mega-constellation satellite
  networks with optical intersatellite links,'' \emph{IEEE Trans Aerosp
  Electron Syst}, vol.~57, no.~6, pp. 4213--4226, Dec. 2021.

\bibitem{MH2003}
M.~Mohorcic, A.~Svigelj, G.~Kandus \emph{et~al.}, ``Demographically weighted
  traffic flow models for adaptive routing in packet-switched non-geostationary
  satellite meshed networks,'' \emph{Comput. Netw.}, vol.~43, no.~2, pp.
  113--131, Oct. 2003.

\bibitem{MH2021}
Q.~Chen, L.~Yang, X.~Liu \emph{et~al.}, ``Multiple gateway placement in
  large-scale constellation networks with inter-satellite links,'' \emph{Int.
  J. Satell. Commun. Netw.}, vol.~39, no.~1, pp. 47--64, Jan. 2021.

\bibitem{GraphMin}
J.~Tao, Z.~Na, B.~Lin \emph{et~al.}, ``A joint minimum hop and earliest arrival
  routing algorithm for leo satellite networks,'' \emph{IEEE Trans. Veh.
  Technol.}, pp. 1--13, Jan. 2023.

\bibitem{Zhu}
J.~Zhu \emph{et~al.}, ``Timing advance estimation in low earth orbit satellite
  networks,'' \emph{IEEE Trans. Veh. Technol.}, 2023, doi:
  10.1109/TVT.2023.3325328.

\bibitem{help2}
X.~Zhang, J.~Wang, C.~Jiang \emph{et~al.}, ``Robust beamforming for multibeam
  satellite communication in the face of phase perturbations,'' \emph{IEEE
  Trans. Veh. Technol.}, vol.~68, no.~3, pp. 3043--3047, Jan. 2019.

\bibitem{yuan2024}
S.~Yuan \emph{et~al.}, ``Joint beam direction control and radio resource
  allocation in dynamic multi-beam leo satellite networks,'' \emph{IEEE Trans.
  Veh. Technol.}, pp. 1--15, 2024, doi: 10.1109/TVT.2024.3353339.

\bibitem{DIS2022}
G.~Stock, J.~A. Fraire, and H.~Hermanns, ``Distributed on-demand routing for
  leo mega-constellations: A starlink case study,'' in \emph{2022 11th Advanced
  Satellite Multimedia Systems Conference and the 17th Signal Processing for
  Space Communications Workshop (ASMS/SPSC)}, Graz, Austria, Sep. 2022, pp.
  1--8.

\bibitem{help3}
C.~Li, W.~He, H.~Yao \emph{et~al.}, ``Knowledge graph aided network
  representation and routing algorithm for leo satellite networks,'' \emph{IEEE
  Trans. Veh. Technol.}, vol.~72, no.~4, pp. 5195--5207, Nov. 2022.
\end{thebibliography}

\vspace{-6 mm}
\begin{IEEEbiography}[{\includegraphics[width=0.98in,clip,keepaspectratio]{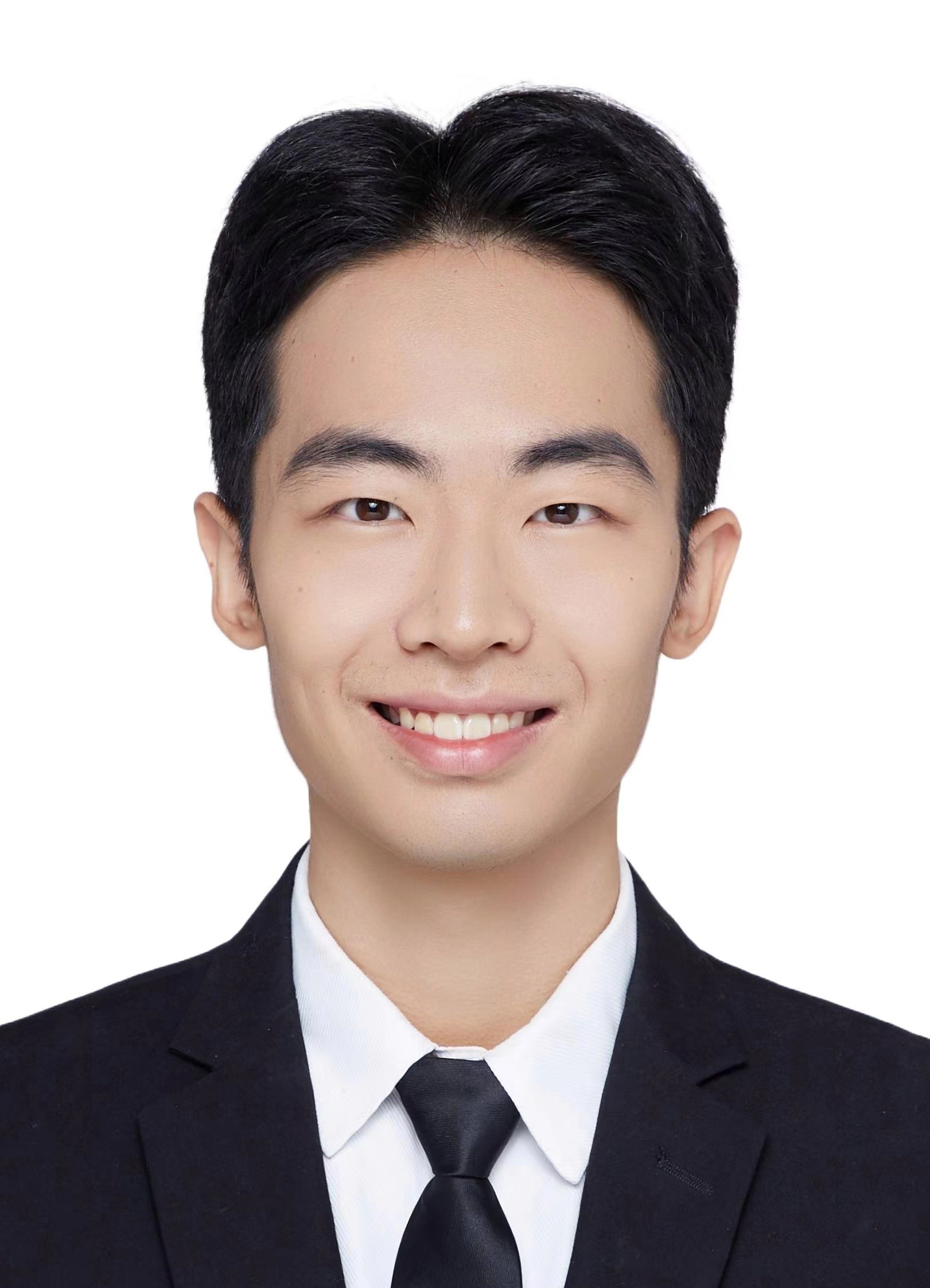}}]
{Xin'ao Feng} received the bachelor's degree in information and communication engineering from Beijing University of Posts and Telecommunications, Beijing, China, in 2022. He is currently working toward a Ph.D degree in the State Key Laboratory of Networking and Switching Technology, Beijing University of Posts and Telecommunications, Beijing, China. His research interests include routing technologies,  LEO satellite communications and mega-constellation networks.
\end{IEEEbiography}
\vspace{-3 mm}
\begin{IEEEbiography}[{\includegraphics[width=1in,height=1.25in,clip,keepaspectratio]{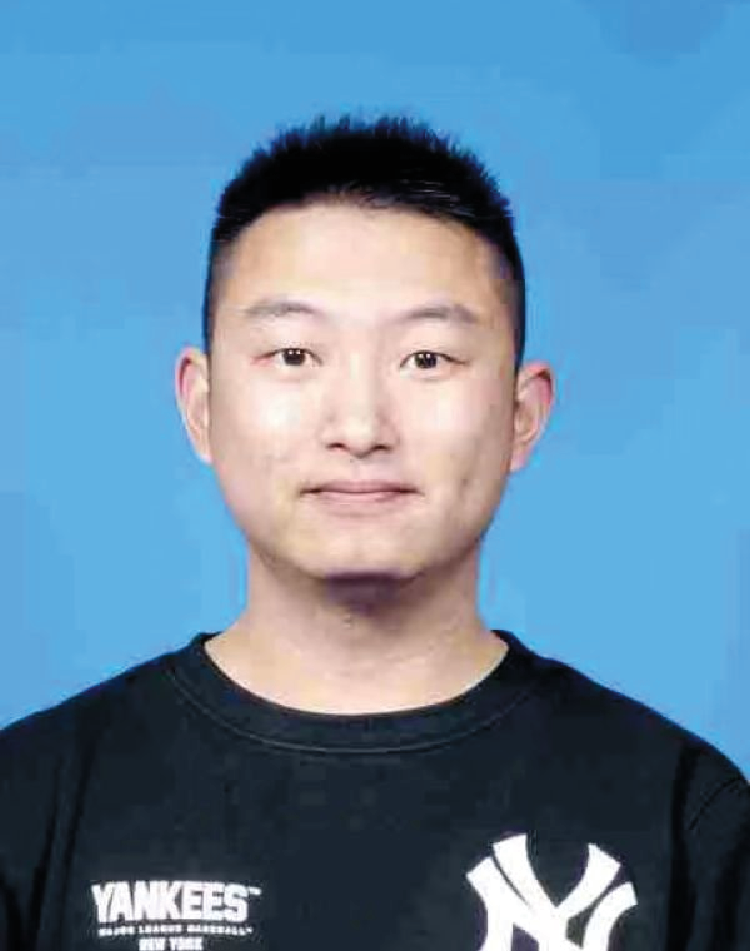}}]
{Yaohua Sun} received the bachelor's degree (Hons.) in telecommunications engineering (with management) and the Ph.D. degree in communication engineering from Beijing University of Posts and Telecommunications (BUPT), Beijing, China, in 2014 and 2019, respectively.

He is currently an Associate Professor with the School of Information and Communication Engineering, BUPT.
His research interests include intelligent radio access networks and LEO satellite communication.
He has published over 30 papers including 3 ESI highly cited papers. He has been a Reviewer for \emph{IEEE Trans. Commun.}, \emph{IEEE Trans. Mob. Comput.}, etc.
\end{IEEEbiography}
\vspace{-3 mm}
\begin{IEEEbiography}[{\includegraphics[width=1in,height=1.25in,clip,keepaspectratio]{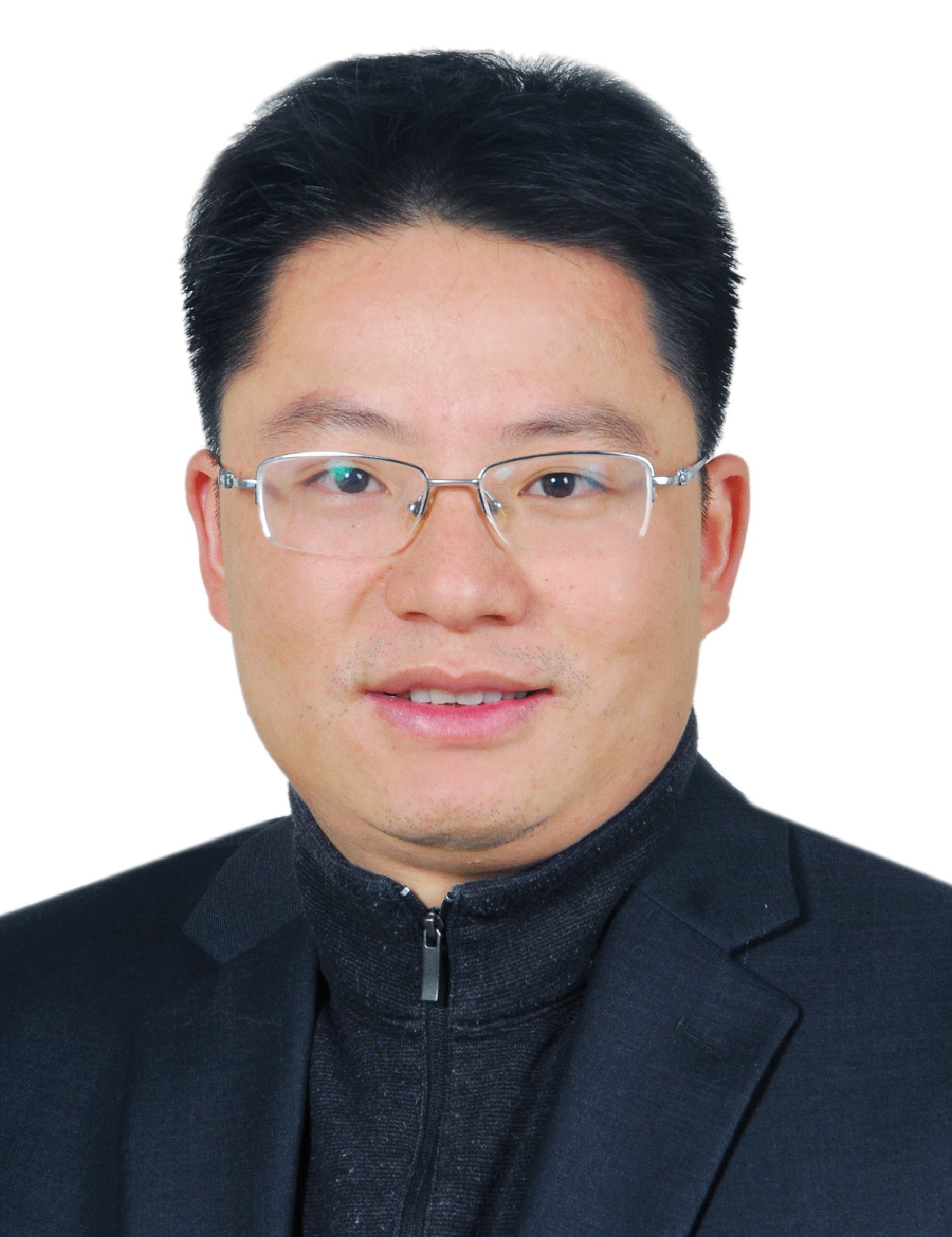}}]
  {Mugen Peng} (Fellow, IEEE) received the Ph.D. degree in communication and information systems from the Beijing University of Posts and Telecommunications, Beijing, China, in 2005. In 2014, he was an Academic Visiting Fellow at Princeton University, Princeton, NJ, USA.
  He joined BUPT, where he has been the Dean of the School of Information and Communication Engineering since June 2020, and the Deputy Director of the State Key Laboratory of Networking and Switching Technology since October 2018.
  He leads a Research Group focusing on wireless transmission and networking technologies with the State Key Laboratory of Networking and Switching Technology, BUPT.
  His main research interests include wireless communication theory, radio signal processing, cooperative communication, self-organization networking, non-terrestrial network, and Internet of Things.
  He was a recipient of the 2018 Heinrich Hertz Prize Paper Award, the 2014 IEEE ComSoc AP Outstanding Young Researcher Award, and the  Best Paper Award in IEEE ICC 2022, JCN 2016, and IEEE WCNC 2015.
  He is/was on the Editorial or Associate Editorial Board of \emph{IEEE Commun. Mag.}, \emph{IEEE Netw.}, \emph{IEEE Internet Things J.}, \emph{IEEE Trans. Veh. Technol.}, and \emph{IEEE Trans. Netw. Sci. Eng.}, etc.
\end{IEEEbiography}
\end{document}